\documentclass[3p,12pt]{elsarticle}
\usepackage{booktabs}
\usepackage[fleqn]{amsmath}
\usepackage{cases}
\usepackage{multirow}
\usepackage{xcolor}
\usepackage[normalem]{ulem}
\usepackage{tabularx}
\usepackage{siunitx}
\usepackage{comment}
\usepackage{algorithm}
\usepackage{algpseudocode}
\usepackage{algorithmicx}
\usepackage{grffile}
\usepackage{rotating}
\usepackage{lineno}
\usepackage{hyperref}
\usepackage{graphicx,enumerate,subfigure}
\usepackage{amssymb}
\usepackage{bm,amsmath}
\usepackage{caption,color,xcolor}
\usepackage{subeqnarray}
\usepackage{multirow}
\usepackage{lscape}
\usepackage{rotating}
\usepackage{graphics}
\usepackage{float}
\usepackage{epstopdf,epsfig}
\usepackage{threeparttable}
\usepackage{soul}
\usepackage{enumitem}

\usepackage[capitalize]{cleveref}

\floatname{algorithm}{Algorithm}
\biboptions{numbers,sort&compress} 
\journal{XXX}

\newcounter{bla}

\begin{document}

\begin{frontmatter}
\title{A fast-converging and asymptotic-preserving method for adjoint shape optimization of rarefied gas flows}

  \author{Yanbing Zhang\corref{equal_contribution}}
  \author{Ruifeng Yuan\corref{equal_contribution}}
  \author{Lei Wu\corref{mycorrespondingauthor}}
  \cortext[equal_contribution]{Both authors contribute equally.}
  \cortext[mycorrespondingauthor]{Corresponding authors: {wul@sustech.edu.cn}}

\address{Department of Mechanics and Aerospace Engineering, Southern University of Science and Technology, Shenzhen 518055, China}

\begin{abstract}
Adjoint-based shape optimization is a powerful technique in fluid-dynamics optimization, capable of identifying an optimal shape within only dozens of design iterations. However, when extended to rarefied gas flows, the computational cost becomes enormous because both the six-dimensional primal and adjoint Boltzmann equations must be solved for each candidate shape. Building on the general synthetic iterative scheme (GSIS) for solving the primal Boltzmann model equation, this paper presents a fast-converging and asymptotic-preserving method for solving the adjoint kinetic equation.
The GSIS accelerates the convergence of the adjoint kinetic equation by incorporating solutions of macroscopic synthetic equations, whose constitutive relations include the Newtonian stress law along with higher-order terms capturing rarefaction effects. As a result, the method achieves asymptotic preservation—allowing the use of large spatial cell sizes in the continuum limit—while maintaining accuracy in highly rarefied regimes.
Numerical tests demonstrate exceptional performance on drag-minimization problems for 3D bodies, achieving drag reductions of 34.5\% in the transition regime and 61.1\% in the slip-flow regime within roughly ten optimization iterations. For each candidate shape, converged solutions of the primal and adjoint Boltzmann equation are obtained with only a few dozen updates of the velocity distribution function, dramatically reducing computational cost compared with conventional methods.
\end{abstract}


\end{frontmatter}

\section{Introduction}

Recent years have seen a growth in engineering applications for rarefied gas flows across many fields, including atmospheric re-entry vehicles \cite{reed2010investigation,li2021kinetic}, vacuum pumps \cite{hablanian1997high,sharipov2005numerical}, extreme ultraviolet lithography \cite{bakshi2009euv}, and nuclear fusion \cite{tantos2020deterministic}. Unlike continuum flows, rarefied gas flows cannot be accurately described by the Navier-Stokes (NS) equations and no-slip boundary condition. Therefore, these growing engineering applications have created an urgent need for efficient optimization methodologies applicable to rarefied gas dynamics, where conventional NS-based methods prove inaccurate.

Within computational design optimization, two principal methodologies dominate the landscape: derivative-free surrogate-based methods \cite{leifsson2015aerodynamic} and gradient-based adjoint optimization methods \cite{jameson1988aerodynamic}. Surrogate-based approaches construct simplified models (such as the Kriging model \cite{han2012hierarchical} or neural networks \cite{sun2019review}) that approximate the relationship between design parameters and objectives. These methods offer implementation simplicity as they bypass derivative calculations, but require numerous objective computations that become prohibitively expensive for complex physics models or large design variable numbers. In contrast, adjoint-based methods compute exact gradients with computational cost independent of design dimension, enabling rapid convergence to optimal solutions even with hundreds of design variables. However, they demand sophisticated mathematical formulation and implementation of the adjoint system.

The computational challenges of rarefied gas dynamics further complicate design optimization. The rarefied gas solvers, whether solving the Boltzmann equation with discrete velocity methods (DVM) or employing the direct simulation Monte Carlo method \cite{Yang1995Rarefied, mieussens2000, bird1994molecular}, impose computational costs orders of magnitude higher than the NS solvers. This computational burden is particularly acute when integrated with optimization frameworks that require multiple flow evaluations (i.e., objective computations). Consequently, incorporating rarefied gas flow solvers directly into surrogate-based optimization methods often leads to unacceptable computational expenses, especially for 3D design problems.

These considerations highlight the compelling need for adjoint-based optimization methods for rarefied gas flows. Utilizing exact gradient information, such methods could dramatically reduce the number of required flow simulations, making design optimization feasible even with expensive rarefied flow solvers \cite{sato2019topology, guan2023densityIPDSMC, guan2024adjointDVM, yuan2024design, yuan2025adjoint}. On this point, the previous work \cite{yuan2025adjoint} has established a foundation for shape optimization in rarefied flows using the Bhatnagar-Gross-Krook (BGK) model equation \cite{Bhatnagar1954} with diffuse-reflection boundary conditions. This approach achieves convergence to a local optimum within roughly ten optimization iterations across the full range of gas rarefaction, demonstrating strong potential for engineering applications---especially when the general shape of the object is known---while offering advantages over topology-optimization methods \cite{sato2019topology, guan2023densityIPDSMC, guan2024adjointDVM, yuan2024design}.
In each optimization iteration, the primal and adjoint kinetic models defined in six-dimensional phase-space must be solved for 3D flows. When the Knudsen number (i.e., the ratio of the molecular mean free path to a reference flow length) is large, these kinetic equations can be solved efficiently. However, in near-continuum flow regimes, conventional numerical schemes require a very large number of iterations and suffer from substantial numerical dissipation when the spatial cell size is not sufficiently small. Thus, an efficient numerical scheme for solving the kinetic equations across the continuum to free-molecular flow regimes is urgently needed.

The general synthetic iterative scheme (GSIS) is one such method capable of solving the primal kinetic equation both efficiently and accurately \cite{su2020can,su2020fast,zhang2024efficient}. In this paper, we introduce significant improvements to the previous rarefied-gas shape-optimization framework \cite{yuan2025adjoint} by developing a GSIS formulation  specifically tailored for the adjoint kinetic equation. This GSIS solver accelerates the convergence of both the primal and adjoint kinetic equations by incorporating macroscopic NS and adjoint-NS equations, thereby delivering excellent computational efficiency and asymptotic-preserving behavior (i.e., the continuum flow dynamics can be accurately recovered even when the cell size is much larger than the molecular mean free path). Across the entire range of gas rarefaction, converged solutions are obtained with only a few dozen updates of the velocity distribution function. This advancement greatly enhances the overall efficiency of the optimization method in rarefied gas dynamics.

The remainder of this paper is organized as follows.
Section \ref{sec:2} establishes the theoretical foundation for adjoint-based shape optimization in rarefied gas flows. Section \ref{sec:gsis} describes the GSIS solver for the adjoint kinetic equations and the associated sensitivity analysis with respect to boundary geometry. Section \ref{sec:meshdeform} introduces the adopted boundary parameterization and mesh deformation techniques. Section \ref{sec:num_example} presents validation studies and optimization results for 3D drag reduction  across different flow regimes. Finally, Section \ref{sec:conclusion} summarizes the main findings.

\section{Formulation}\label{sec:2}

In this section, we first establish the theoretical foundation for the adjoint-based shape-optimization in rarefied gas flows, including the underlying gas-kinetic theory and the associated boundary conditions. Second, we formulate the steady-state shape-optimization problem by specifying the objective functional and geometric constraints. Third, we present the sensitivity analysis and derive the corresponding adjoint equations. Finally, we outline the overall workflow of shape optimization in rarefied gas flows.

\subsection{Gas-kinetic theory}

In gas kinetic theory, the state of gas is characterized by the velocity distribution function $f(t,\bm{x},\bm{v})$, where $t$ is the time, $\bm{x}=(x,y,z)$ is the spatial Cartesian coordinates, and  $\bm{v}=(v_x,v_y,v_z)$ is the molecular velocity space. Macroscopic flow variables are obtained by integrating $f$ over the molecular velocity space. Specifically, the conservative variables are given by
\begin{equation}\label{eqn:w_int}
\bm W = \int {\bm \psi fd\Xi },
\end{equation}
where $\bm W=(\rho,\rho\bm u,\rho E)^\top$ is the vector for the densities of mass, momentum, and energy, with $E={u}^2/2 + {RT}/({\gamma  - 1})$, $T$ being the gas temperature, $R$ the specific gas constant, and $\gamma$ the specific heat ratio.
The moment vector $\bm \psi$ and the velocity-space element are defined as
\begin{equation}
\bm{\psi} = \left(1,\, \bm{v},\, \frac{v^2}{2}  \right)^\top, \quad d\Xi = dv_x dv_y dv_z,
\end{equation}
where $\top$ denotes the matrix transpose.

The evolution of velocity distribution function is governed by the Boltzmann equation~\cite{Cercignanibook1988}. However, due to the high computational cost associated with its collision operator, we adopt the widely used Bhatnagar--Gross--Krook (BGK) model equation as the primal kinetic equation~\cite{Bhatnagar1954}:
\begin{equation}\label{eqn:bgk_org}
\begin{aligned}
\frac{{\partial f}}{{\partial t}}{\rm{ + }}\bm v \cdot \nabla f = \frac{{g - f}}{\tau }, \\
\text{with}~
g =g_{\rm M}(\rho,\bm{u},T)= \frac{\rho}{(2\pi RT)^{3/2}}
{\exp\left( - \frac{{{{(\bm v - \bm u)}^2}}}{{2RT}}\right)},
\end{aligned}
\end{equation}
where $\nabla$ denotes the spatial gradient operator, $g$ represents the local equilibrium  distribution, and $\tau$ is the relaxation time. 
The latter is related to the dynamic viscosity $\mu$ and gas pressure $p = \rho R T$ as
\begin{equation}
\tau  = \frac{\mu}{p}.
\end{equation}
For simplicity, we assume a hard-sphere molecular model, for which $ \mu \propto \sqrt{T} $.


The degree of gas rarefaction is quantified by the Knudsen number Kn, defined as the ratio of the molecular mean free path to a characteristic length scale $ l_{\rm ref} $:
\begin{equation}\label{eqn:kndefine}
{\rm{Kn}} = \frac{\tau }{{{l_{{\rm{ref}}}}}}\sqrt {\frac{\pi RT}{2 }},
\end{equation}
based on which the gas flows can be categorized into four regimes~\cite{tsien1946superaerodynamics}. In the continuum regime ($\text{Kn} \lessapprox 0.001$), the NS equations with no-velocity slip and no-temperature-jump boundary conditions remain valid. In the slip regime ($0.001 \lessapprox \text{Kn} \lessapprox 0.1$), velocity slip and temperature jump occur at solid boundaries. As Kn increases further, the transition regime ($0.1 \lessapprox \text{Kn} \lessapprox 10$) arises, where the NS equations fail and gas behavior must be described by molecular transport–collision dynamics. Finally, in the free-molecular regime ($\text{Kn} \gtrapprox 10$), the flow is dominated by free molecular transport with negligible intermolecular collisions.

It should be noted that, while the  kinetic equation is valid across all flow regimes, it becomes stiff in the continuum limit due to rapid collisional relaxation. This necessitates specialized numerical treatment, which will be addressed in Section~\ref{sec:gsis}.

\subsection{Kinetic boundary conditions}\label{sec:boundary}

The kinetic equation describes gas-gas interactions. To fully characterize rarefied gas flows in wall-confined configurations, appropriate gas-surface boundary conditions must be specified. In this study, two types of boundary conditions are considered.

The first is the Dirichlet boundary condition applied at the far-field boundary $\Gamma _{\rm d}$.  For incoming gas molecules---i.e., those with molecular velocity $\bm v$ satisfying $\bm v \cdot \bm n < 0$, where $\bm n$ is the outward-pointing unit normal vector---the distribution function is prescribed as
\begin{equation}\label{eqn:bc_drlt}
f = f_{\rm d} \quad {\rm{in}}\quad {\Gamma _{\rm d}} \times {\Xi ^ - }.
\end{equation}
where $f_{\rm d}$ is a given distribution (normally a Maxwellian distribution corresponding to the free-stream gas state). Here, ${\Xi ^ \pm } = \left\{ \bm v | (\bm v \cdot \bm n) \gtrless 0 \right\}$ denotes the set of velocities for molecules exiting/entering the computational domain.

The second is the diffuse-reflection boundary condition imposed on the gas--solid interface ${\Gamma _{\rm w}}$. When gas molecules strike the solid wall, they are re-emitted with a Maxwellian velocity distribution:
\begin{equation}\label{eqn:formula_bgk_fdw0}
f = {g_{\rm{M}}}({\rho _{\rm{w}}},{\bm u_{\rm{w}}},T_{\rm{w}}),
\quad {\rm{in}}\quad {\Gamma _{\rm w}} \times {\Xi ^ - },
\end{equation}
where $\bm u_{\rm{w}} = \bm 0 $ (since the wall is stationary), $T_{\rm{w}}$ is the prescribed wall temperature, and $\rho _{\rm{w}}$ is determined by enforcing zero net mass flux across the boundary:
\begin{equation}\label{eqn:formula_bgk_fdw}
\int_{{\Xi ^ + }} {(\bm v \cdot \bm n){f}d\Xi } + \int_{{\Xi ^ - }} {(\bm v \cdot \bm n){{g_{\rm{M}}}({\rho _{\rm{w}}},{\bm 0},T_{\rm{w}})}d\Xi }  = 0.
\end{equation}
This diffuse boundary condition naturally recovers the no-slip condition in the continuum regime ($\text{Kn} \lessapprox 0.001$), while correctly capturing non-equilibrium effects---such as velocity slip and temperature jump---in rarefied gas flows. Consequently, it provides a unified treatment across the entire range of Knudsen number.

\subsection{Optimization problem}

Consider a solid body immersed in a gas flow domain $\Omega$, which is bounded by a far-field boundary $\Gamma _{\rm d}$ and a gas--solid interface $\Gamma _{\rm w}$. The goal is to optimize the shape of $\Gamma _{\rm w}$, which is parameterized by a set of design variables (to be detailed in Section~\ref{sec:gsis}). We restrict our attention to steady-state flows, so all quantities are time-independent.

The objective functional $J$ is chosen to represent a physical quantity of interest, typically the total force or heat flux transferred across the gas-solid boundary and exerted on the solid body. It can be expressed as an integral over $\Gamma _{\rm w}$ of a specific moment of the incoming mass flux $(\bm v \cdot \bm n) f$:
\begin{equation}\label{eqn:formula_optJ}
J = \int_{{\Gamma _{\rm{w}}}} {\int_\Xi  {m_J(\bm v \cdot \bm n){f}} d\Xi d\Gamma } , 
\end{equation}
where $m_J$ specifies the moment corresponding to the target quantity. For instances, $ m_J =v_x$ for the force in the x direction, or $ m_J =v^2/2$ for the heat flux. 
To ensure physical feasibility, a volume constraint is imposed on the solid body. The volume, computed using the Gauss divergence theorem, must remain no less than a prescribed minimum $V_{\min}$:
\begin{equation}
\label{eq:volume-constraint}
G = \frac{1}{3}\int_{{\Gamma _{\rm{w}}}} {\bm x \cdot \bm nd\Gamma }  + {V_{\min }} \le 0.
\end{equation}
Thus, the complete optimization problem reads:
\begin{equation}\label{eqn:formula_opt}
\mathop {\min }\limits_{\Gamma _{\rm w}}  ~J, 
\quad {\rm s.t.}\quad G \le 0.
\end{equation}

Note that the distribution function $f$ is not a design variable; rather, it is implicitly determined by the geometry of $\Gamma _{\rm w}$ through the steady-state gas-kinetic system:
\begin{equation}\label{eqn:formula_bgk}
\left.
\begin{aligned}
\bm v \cdot \nabla f - \frac{{{ g} -  f}}{{{\tau}}} &= 0 \quad {\rm{in}}\quad \Omega  \times \Xi, \\
 f - { f_{\rm d}} &= 0\quad  {\rm{in}}\quad {\Gamma _{\rm d}} \times {\Xi ^ - }, \\
 f - {g_{\rm{M}}}({\rho _{\rm{w}}},{\bm 0},T_{\rm{w}}) &= 0\quad  {\rm{in}}\quad {\Gamma _{\rm w}} \times {\Xi ^ - },
\end{aligned}
\right\}
\end{equation}
where the boundary conditions are defined in Section~\ref{sec:boundary}. This nested structure, where the state $f$ depends on the design ${\Gamma _{\rm w}}$, necessitates adjoint-based sensitivity analysis for efficient gradient computation, which is developed in the next section.

\subsection{Sensitivity analysis}\label{sec:sens}

To solve the optimization problem~\eqref{eqn:formula_opt}, a gradient-based approach is adopted, which requires the computation of the sensitivity, i.e., the derivative of the objective functional $J$  with respect to the geometry of the solid boundary $\Gamma _{\rm w}$. This sensitivity is efficiently obtained through adjoint analysis. Specifically, the continuous adjoint formulation is first employed for its conceptual clarity and ease of derivation, while the final sensitivity is computed using a hybrid strategy that combines elements of both the continuous and discrete adjoint methods. The rationale for adopting this hybrid strategy has been discussed in detail in Refs.~\cite{yuan2024design,yuan2025adjoint}.

Recall that the objective functional $J$ has dependence on the distribution function $f$, which itself is governed by the kinetic equation~\eqref{eqn:formula_bgk} and depends implicitly on the geometry of $\Gamma _{\rm w}$. To account for this dependency, we introduce the following Lagrangian with the Lagrange multipliers (i.e., adjoint variables $\phi ,{\varphi _{\rm{d}}},{\varphi _{\rm{w}}}$)  as:
\begin{equation}\label{eqn:lagrangian}
{\cal L}(f,\theta,\phi,\varphi _{\rm{d}},\varphi _{\rm{w}} ) =  J + I + B_{\rm w} + B_{\rm d},
\end{equation}
where
\begin{equation}
    \begin{aligned}
        I=&\int_\Omega  {\int_\Xi  {\phi   \left( {\bm v \cdot \nabla f - \frac{{{ g} -  f}}{{{\tau}}}} \right)d\Xi } d\Omega },\\
      B_{\rm w}=&\int_{{\Gamma _{\rm w}}} {\int_{{\Xi ^ - }} {{ \varphi _{\rm w}}  \left[ { f - {g_{\rm{M}}}({\rho _{\rm{w}}},{\bm u_{\rm{w}}},T_{\rm{w}})} \right]d\Xi } d\Gamma },\\
      B_{\rm d}=&\int_{{\Gamma _{\rm d}}} {\int_{{\Xi ^ - }} {{ \varphi _{\rm d}} \left( { f - { f_{\rm d}}} \right)d\Xi } d\Gamma }.
    \end{aligned}
\end{equation}

The key idea of computing the total derivative $dJ/d\Gamma _{\rm w}$
is to choose the adjoint variables such that the variation of $\cal L$ with respect to perturbations in $f$ vanishes, i.e., $d{\cal L}(f;\delta f) = 0$. This eliminates the indirect dependence of $J$ on $\Gamma _{\rm w}$ through $f$, allowing the total derivative $dJ/d\Gamma _{\rm w}$ to be evaluated solely from the explicit geometric dependence in ${\cal L}$ (since $J \equiv {\cal L}$). Carrying out the variational analysis and integrating by parts, we obtain the adjoint kinetic system:
\begin{equation}\label{eqn:formula_adjointbgk}
\left.
\begin{aligned}
 - \bm v \cdot \nabla  \phi  = \frac{{{ \phi _{{\rm{eq}}}} -  \phi }}{{{\tau }}} + { \phi _\tau }, \quad {\rm{in}}\quad \Omega  \times \Xi, \\
 \phi  =  0, \quad {\rm{in}} \quad {\Gamma _{\rm d}} \times {\Xi ^ + }, \\
 \varphi _{\rm{w}}  = - (\bm v \cdot \bm n)\left( {{\phi } + {m_J}} \right),\quad {\rm{in}}\quad {\Gamma _{\rm w}} \times {\Xi ^ - }, \\
 \phi  = \sqrt {\frac{{2\pi }}{{R{T_{\rm{w}}}}}} \int_{{\Xi ^ - }} \varphi _{\rm{w}}   {g_{\rm{M}}}(1,\bm 0,T_{\rm{w}})d\Xi -{m_J} ,\quad {\rm{in}}\quad {\Gamma _{\rm w}} \times {\Xi ^ + },
\end{aligned}
\right\}
\end{equation}
where the target distribution function ${ \phi _{\rm eq}}$ and source term ${ \phi _\tau }$ are defined as
\begin{equation}\label{eqn:adjoint_eqdefines}
{ \phi _{\rm eq}} = \hat {\bm W} \cdot \bm{\psi},\quad
{ \phi _\tau } =  - \frac{1}{{{\tau }}}\frac{{\partial {\tau }}}{{\partial \bm W}} \cdot \bm{\psi} \int_\Xi  { \phi   \frac{{{ g } -  f}}{{{\tau }}}d\Xi },
\end{equation}
with the adjoint macroscopic variable (which is analogous to the macroscopic variable $\bm W$ in Eq.~\eqref{eqn:w_int}, and the moment weight $\partial {g }/\partial {\bm W }$ in the definition of $\hat {\bm W}$ plays a role analogous to the moment vector $\bm{\psi}$ in the primal system) defined as 
\begin{equation}\label{eqn:adjoint_macdefines}
\hat {\bm W} = \int_\Xi  { \phi   \frac{{\partial { g }}}{{\partial {\bm W }}}d\Xi } . 
\end{equation}
 Moreover, the objective-related term $m_J$ appears as a boundary source in the adjoint equation, reflecting its influence on the sensitivity.

Once the adjoint variables---particularly $\varphi _{\rm{w}}$---are solved, the total sensitivity of $J$ with respect to shape variations $\delta \Gamma _{\rm w}$ is obtained from the explicit geometric dependence in the Lagrangian:
\begin{equation}\label{eqn:sens_shape}
d{\cal L}(\Gamma _{\rm w} ;\delta \Gamma _{\rm w} ) = dJ(\Gamma _{\rm w} ;\delta \Gamma _{\rm w} )+dB_{\rm w}(\Gamma _{\rm w};\delta \Gamma _{\rm w} ), 
\end{equation}
since the volume integral $I$ and far-field boundary term $B_{\rm d}$ are unaffected by perturbations of $\Gamma _{\rm w}$ (as $f$ is held fixed during this variation).

In practice, the primal and adjoint equations are solved separately by means of GSIS (see Section~\ref{sec:scheme}). After the solutions for  $f$ and $\varphi _{\rm{w}}$ are obtained, the sensitivity on the discrete boundary $\Gamma _{\rm w}$  is evaluated using an approach analogous to the discrete adjoint method, as detailed in Section~\ref{sec:sens_nod}.

\subsection{Overall workflow}

\begin{figure}[h]
 	\centering
 	\includegraphics[width=0.7\textwidth]{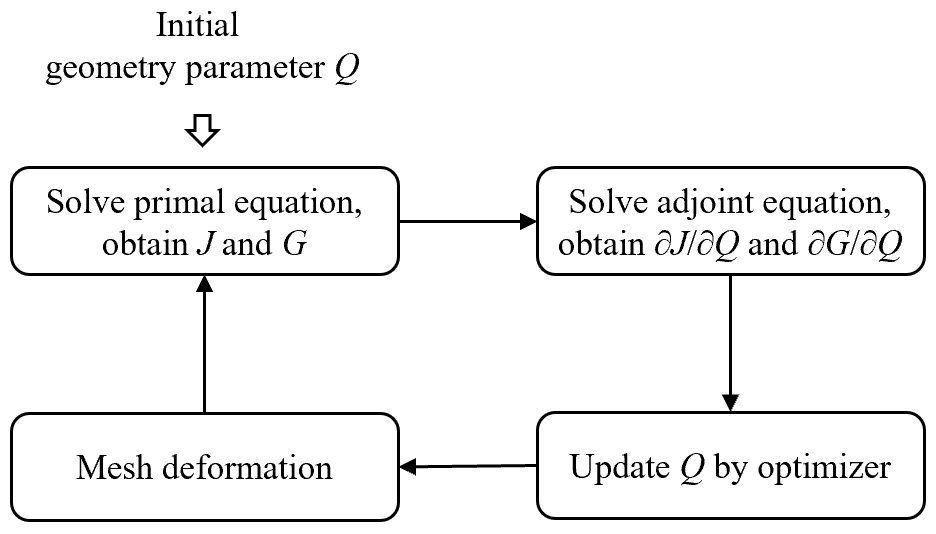}
 	\caption{\label{fig:flowchart} 
    General procedure for optimizing rarefied gas flows.}
 \end{figure}

The proposed shape optimization method is carried out following the standard workflow of gradient-based optimization, and the general computational procedure is illustrated in Fig.~\ref{fig:flowchart}. The boundary-geometry parameterization and the mesh deformation are discussed later in Section~\ref{sec:meshdeform}. The numerical scheme for the adjoint equation is introduced in Section~\ref{sec:scheme}. The method of moving asymptotes \cite{svanberg1987method,svanberg2002class} is employed as the optimizer, as the volume constraints is naturally accommodated through the incorporation of both the value of $G$ and its sensitivity into the algorithm. The NLopt library by Johnson \cite{johnson2007nLopt} is used for implementation, while the computation of $G$ and its sensitivity is straightforward and thus omitted.”
 
\section{Numerical methods for the adjoint kinetic equation and shape sensitivity}\label{sec:gsis}

In this section, we first extend the GSIS to solve the adjoint equation accurately and efficiently. We then introduce the detailed algorithm for sensitivity analysis with respect to the boundary mesh nodes. 

\subsection{Numerical schemes for the adjoint kinetic equation}\label{sec:scheme}

The Boltzmann equation, defined in the six-dimensional phase space, is notoriously difficult to solve numerically. This is because, first, the numerical scheme must converge rapidly to the steady state, and second, it should asymptotically preserve the NS limit in near-continuum flows so that the spatial grid resolution can be significantly reduced. Recently, this two problems are tackled in the GSIS~\cite{su2020can,su2020fast,zhang2024efficient}. Therefore, in this section, we shall focus on the development of GSIS for the adjoint kinetic equation.

The adjoint kinetic equation \eqref{eqn:formula_adjointbgk} is discretized by the finite-volume discrete velocity method with second-order accuracy in physical space:
\begin{equation}\label{eqn:numerad_gov}
 - \sum\limits_{j \in N\left( i \right)} {{A_{ij}}{\bm v_k} \cdot {\bm n_{ij}} \phi _{ij,k}}  =  {V_i}\frac{{\phi _{{\rm eq},i,k} -  \phi _{i,k}}}{{\tau _{i}}} + {V_i} \phi _{\tau ,i,k} ,
\end{equation}
where the subscripts $i,k$ correspond to the discretizations in physical
space and velocity space, respectively; $j$ denotes the neighboring cell of cell $i$ and $N\left( i \right)$ is the set of all neighbors of $i$; $ij$ denotes the variable at the interface between the cell $i$ and $j$; $A_{ij}$ is the interface area, ${\bm n_{ij}}$ is the outward normal unit vector of the interface $ij$ relative to the cell $i$, and ${V_i}$ is the control volume of the cell $i$.

Implicit pseudo time-marching method is used to solve the discretized adjoint kinetic equation. Specifically, if a conventional DVM is used, the temporal discretization for the adjoint kinetic equation will be like:
\begin{equation}\label{eqn:implicit_kinetic0}
\frac{{\phi _i^{n + 1} - \phi _i^n}}{{\Delta {t_{\rm psu}}}} - \frac{1}{{{V_i}}}\sum\limits_{j \in N(i)} {{{\bm v}_k} \cdot {\bm n_{ij}}\phi _{ij,k}^{n + 1}}  = {V_i}\frac{{\phi _{{\rm eq},i,k}^n - \phi _{i,k}^{n + 1}}}{{{\tau _i}}} + {V_i}\phi _{\tau ,i,k}^n,
\end{equation}
where the $\Delta t_{\rm psu}$ is the pseudo time step, and $n$ is the iteration step. When Kn is small, like the primal kinetic equation, the iteration step can be huge in the continuum limit \cite{wang2018comparative}.

 \begin{figure}[h]
 	\centering
 	\includegraphics[width=0.7\textwidth]{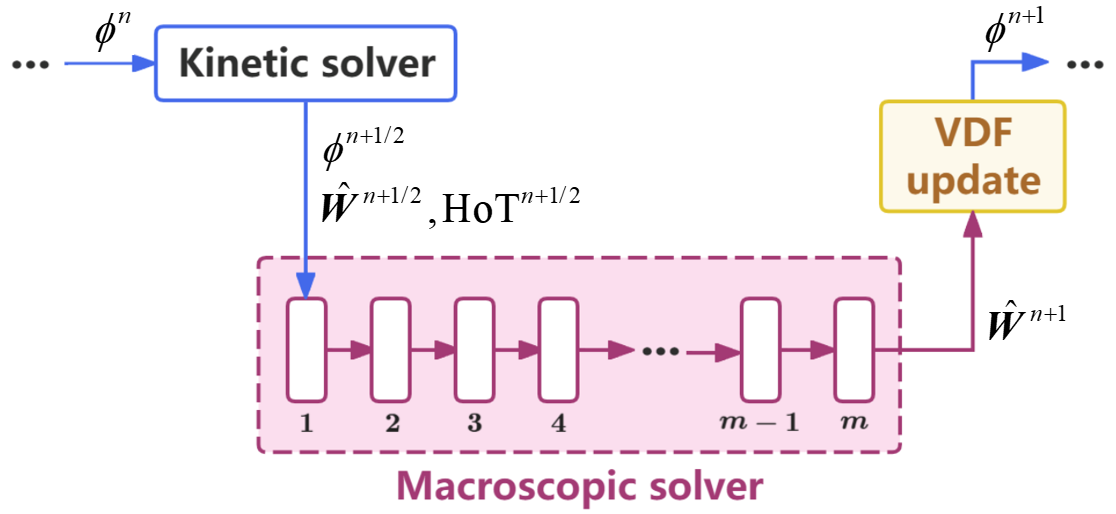}
 	\caption{\label{fig:gsis} Schematic for the GSIS algorithm for the adjoint equation. First, the intermediate velocity distribution function is obtained by solving Eq.~\eqref{eqn:implicit_kinetic0}, when $n+1$ is replaced by $n+1/2$. Second, with the macroscopic adjoint variables and high-order term,  the adjoint-NS-based equation \eqref{eq:adjoint_macro_discreted} is solved to the steady state (or with maximum 1000 iterations of $m$), to get the prediction of adjoint macroscopic quantities $\hat{\bm{W}}$ at the $(n+1)$-th step. Finally, the adjoint distribution function is updated as per Eq.~\eqref{adjoint_vdf_update}.
 }
 \end{figure}

The GSIS~\cite{su2020can,su2020fast,zhang2024efficient} is applied to the adjoint kinetic equation to get the steady-state solution accurately and efficiently. A schematic of the algorithm is shown in Fig.~\ref{fig:gsis}. First, the adjoint kinetic equation is solved using the conventional DVM  for an intermediate $\phi^{n+1/2}$, where $n+1$ in Eq.~\eqref{eqn:implicit_kinetic0} is replaced by $n+1/2$. Then, a macroscopic approximate governing equation, combined with the high-order term (HoT) extracted from the adjoint kinetic system, is solved to obtain the macroscopic state $\hat{\bm W}^{n+1}$. Finally, $\phi^{n+1}$ for the next step can be calculated from the difference between $\hat{\bm W}^{n+1/2}$ and $\hat{\bm W}^{n+1}$. As will be shown in numerical tests, this can effectively accelerate the convergence of the kinetic equation through the solution of the macroscopic equation. The numerical schemes for the kinetic solver and the macroscopic solver are elaborated below.

For the kinetic solver, we solve $\phi _{i,k}^{n + 1/2}$ from Eq.~\eqref{eqn:implicit_kinetic0} through simply replacing  $\phi^{n + 1}$ by $\phi^{n + 1/2}$.
At the cell interface, $\phi _{ij,k}^{n + 1/2}$ is calculated in an upwind manner:
\begin{equation} \label{eq:upwind_interface}
\phi _{ij,k}^{n + 1/2} = \left\{ \begin{aligned}
\phi _{i,k}^{n + 1/2} + \left( {{{\bm x}_{ij}} - {{\bm x}_i}} \right) \cdot \nabla \phi _{i,k}^{n + 1/2},\quad  - {{\bm v}_k} \cdot {{\bm n}_{ij}} > 0\\
\phi _{j,k}^{n + 1/2} + \left( {{{\bm x}_{ij}} - {{\bm x}_j}} \right) \cdot \nabla \phi _{j,k}^{n + 1/2},\quad  - {{\bm v}_k} \cdot {{\bm n}_{ij}} < 0
\end{aligned} \right.,
\end{equation}
where $\nabla \phi _{i,k}^{n + 1/2}$ and $\nabla \phi _{j,k}^{n + 1/2}$ can be calculated by the data reconstruction. By rearranging Eq.~\eqref{eqn:implicit_kinetic0} into an incremental form, and using a first-order upwind scheme to handle the increment of the interfacial $\phi _{i,k}^{n + 1/2}$, we obtain:
\begin{equation}\label{eqn:implicit_kinetic_update}
\begin{aligned}[b]
& \left( {\frac{{V_i}}{{\Delta {t_{\rm psu}}}} + \frac{{V_i}}{{\tau _i}} - \sum\limits_{j \in N_k^ - (i)} {{A_{ij}}{\bm v_k} \cdot {\bm n_{ij}}} } \right)\Delta \phi_{i,k}^{n} \\
=  & V_i \frac{{\phi _{{\rm eq},i,k}^{n} - \phi _{i,k}^n}}{{\tau _{i}}} + V_i \phi _{\tau ,i,k}^{n} + \sum\limits_{j \in N\left( i \right)} {{A_{ij}}{\bm v_k} \cdot {\bm n_{ij}}\bm \phi _{ij,k}^n}  + \sum\limits_{j \in N_k^ + (i)} {{A_{ij}}{\bm v_k} \cdot {\bm n_{ij}}\Delta \phi _{j,k}^{n}}
\end{aligned},
\end{equation}
where $\Delta \phi _{i,k}^n = \phi _{i,k}^{n + 1/2} - \phi _{i,k}^n$ is the increment, $N_k^ + (i)$ is the set of $i$'s neighbors satisfying ${\bm v_k} \cdot {\bm n_{ij}} > 0$, while $N_k^ - (i)$ corresponds to those satisfying ${\bm v_k} \cdot {\bm n_{ij}} < 0$. The above equation is solved using the point relaxation algorithm~\cite{Rogers1995Comparison}. Then the adjoint distribution function $\phi^{n+1/2}$ at the intermediate step can be obtained, and the corresponding adjoint macroscopic variable $\hat{\bm W}^{n+1/2}$ can be calculated through the numerical integration of Eq.~\eqref{eqn:adjoint_macdefines}.

For the macroscopic equation, in this work we solve an adjoint NS system to obtain an approximate macroscopic state $\hat{\bm W}^{n+1}$, which is then used to correct $\phi^{n+1/2}$ and thereby accelerate the convergence of the kinetic iteration. To construct the macroscopic adjoint system, we proceed in a way analogous to the derivation of the primal macroscopic equations from the kinetic system. Starting from the adjoint kinetic equation \eqref{eqn:formula_adjointbgk}, we multiply it by $\partial g / \partial \bm W$ and integrate over the velocity space to obtain the exact moment equation:
\begin{equation}
  -\,
      \int_\Xi \bm{v}\,\cdot \nabla \phi\,
         \frac{\partial g}{\partial \bm W}\,d\Xi
  = \int_\Xi \phi_\tau\,
      \frac{\partial g}{\partial \bm W}\,d\Xi .
\end{equation}
The integral on the left-hand side represents the total adjoint flux generated by the kinetic equation. In order to obtain a tractable macroscopic closure, we introduce an adjoint Euler (convective) flux based on the equilibrium adjoint distribution,
\begin{equation}
  \hat{\bm{F}}^{c}
  = \int_\Xi \bm{v}\,\phi_{\mathrm{eq}}\,
      \frac{\partial g}{\partial \bm W}\,d\Xi = \hat {\bm W} \cdot \frac{{\partial {{\bf{F}}_{\rm{E}}}}}{{\partial \bm W}},
\end{equation}
where ${{\bf{F}}_{\rm{E}}}$ is the flux tensor of the Euler equation defined as ($\bf I_{3}$ is a $3\times3$ identity matrix)
\begin{equation}\label{eqn:numerpr_eulerflux}
{\bf F_{{\rm E}}}(\bm W) = 
\left( 
\begin{aligned}
&\quad\quad \rho \bm u \\
&\rho \bm u \bm u  + p {\bf I_{3}}\\
&\left( {\rho E + p} \right)\bm u 
\end{aligned} 
\right),
\end{equation}
and complement it with a viscous flux $\hat{\bm{F}}^{v}$ whose tensorial form is taken from the laminar continuous adjoint NS operator (\ref{appendix_flux_adjoint}, see also Ref.~\cite{bueno2012continuous}). All remaining moment contributions of the kinetic transport operator and the source term $\phi_\tau$ that are not captured by $\hat{\bm{F}}^{c} + \hat{\bm{F}}^{v}$ are collected into a high-order correction term $\mathbf{HoT}$. With these definitions, the macroscopic adjoint governing equation can be written as
\begin{equation}
 -\nabla \hat {\bm W}:\frac{{\partial {{\bf{F}}_{\rm{E}}}}}{{\partial \bm W}}
     - \nabla \hat{\bm{F}}^{v}
  + \mathbf{HoT} = 0 ,
  \label{eq:adjoint_macro_equation}
\end{equation}
which serves as the macroscopic component of the GSIS-based adjoint solver. The explicit expressions of $\hat{\bm{F}}^{c}$ and $\hat{\bm{F}}^{v}$ are summarized in ~\ref{appendix_flux_adjoint}.

According to the construction of the GSIS method~\cite{su2020can,su2020fast,zhang2024efficient},  $\mathbf{HoT}$, as the difference between the flux integrated from $\phi^{n+1/2}$ of the kinetic system and the flux from the adjoint NS equations, is computed as follows:
\begin{equation}\label{eqn:hot}
\begin{aligned}
{\bf{HoT}}_i^{n + 1/2} = & - \int_\Xi ^{} {\bm v \cdot \nabla \phi _i^{n + 1/2}\frac{{\partial g}}{{\partial {{\bm W}_i}}}d\Xi }  - \int_\Xi ^{} {\phi _{\tau ,i}^{n + 1/2}\frac{{\partial g}}{{\partial {{\bm W}_i}}}d\Xi } \\
& + \frac{1}{{{V_i}}}\sum\limits_{j \in N(i)} {{A_{ij}}\hat {\bm F}_{ij}^{c,n + 1/2}} + \frac{1}{{{V_i}}}\sum\limits_{j \in N(i)} {{A_{ij}}\hat {\bm F}_{ij}^{{v},n + 1/2}},
\end{aligned}
\end{equation}
where the adjoint convective flux \(\hat{\bm{F}}_{ij}^{c}\) and viscous flux \(\hat{\bm{F}}_{ij}^{v}\) are provided in \ref{appendix_flux_adjoint}. 
It should be emphasized that, the gradients $\nabla \phi _i^{n + 1/2}$ and $\nabla \hat{ \bm W}_i^{n + 1/2}$ in Eq.~\eqref{eqn:hot} must be reconstructed using exactly the \emph{same} linear reconstruction operator (ensuring that the superposition principle is satisfied). In this work, we employ a least-squares reconstruction (without limiters). This enables the numerical scheme to possess multiscale properties and yield low-numerical-dissipation solutions in continuum flows, as discussed in Ref.~\cite{yuan2021novel}.



After obtaining the HoT, $\hat{\bm W}^{n+1}$ can be solved from the following adjoint-NS-based equation in the finite-volume framework:
\begin{equation}\label{eq:adjoint_macro_discreted}
    -\frac{1}{V_i}\sum_{j \in N(i)} \left(\hat{\bm{F}}_{ij}^{c, n+1} + \hat{\bm{F}}_{ij}^{v, n+1} \right) A_{ij} + \mathbf{HoT}_i^{n+1/2} = \bm 0.
\end{equation}
To this end, the implicit time-marching scheme is adopted:
\begin{equation}\label{eqn:macsgs}
        \left(\frac{1}{\Delta t_{\rm psu}} + \frac{1}{2 V_i} \sum_{j \in N(i)} \varrho_{ij}^{i} A_{ij}\right) \Delta \hat{\boldsymbol{W}}_{i}^{m} = \boldsymbol{R}_{i}^{m} + \frac{1}{2 V_i} \sum_{j \in N(i)} \left( \Delta \hat{\boldsymbol{W}}_{j}^{m} \cdot \mathbf{J}_{ij}^{i} + \varrho_{ij}^{i} \Delta \hat{\boldsymbol{W}}_{j}^{m}  \right) A_{ij},
\end{equation}
with the residual 
\begin{equation}
        \boldsymbol{R}_{i}^{m} =  \frac{1}{V_i} \sum_{j \in N(i)} \left(\hat{\bm{F}}_{ij}^{c, m} + \hat{\bm{F}}_{ij}^{v, m} \right) A_{ij} -\mathbf{HoT}_i^{n+1/2}.
\end{equation}
Here, $\Delta \hat{\boldsymbol{W}}_i^m = \hat{\boldsymbol{W}}_i^{m+1} - \hat{\boldsymbol{W}}_i^{m}$ is the increment with $m$ corresponding to the inner iteration to solve Eq.~\eqref{eq:adjoint_macro_discreted}, $\Delta t_{\rm psu}$ is the pseudo time step, $\mathbf{J}_{ij}^{i}= \partial ({{\bf{F}}_{\rm{E}}} \cdot {{\bm n}_{ij}})/\partial {{\bm W}_i}$ is the Jacobian matrix of the Euler flux, and $\varrho_{ij}^{i}$ is the spectral radius of the Jacobian, given in \ref{appendix_flux_adjoint}. Note that $\mathbf{HoT}_i^{n+1/2}$ remains unchanged during the iteration of $m$, and Eq.~\eqref{eqn:macsgs} is solved using the point relaxation algorithm.

Finally, once \(\hat{\boldsymbol{W}}^{n+1}\) is obtained, the adjoint distribution function \(\phi^{n+1}\) is updated as:
\begin{equation}\label{adjoint_vdf_update}
    \phi^{n+1} = \phi^{n+1/2} + \left[ \phi_{eq}\left( \hat{\bm{W}}^{n+1} \right) - \phi_{eq}\left( \hat{\bm{W}}^{n+1/2} \right) \right].
\end{equation}




\subsection{Boundary flux configuration for the adjoint NS equations}

Within the GSIS framework, the adjoint NS system is solved in conservative form on the same finite-volume mesh as the kinetic equation. To maintain consistency between the macroscopic and kinetic adjoint solutions at the boundaries and to improve convergence, similar to the consistent boundary condition in the primal kinetic equation \cite{liu2024further}, the adjoint NS boundary flux is constructed and subsequently updated from the kinetic adjoint solution in two steps.

First, at the beginning of each GSIS iteration, an initial value of the adjoint NS boundary flux is obtained directly from the kinetic equation by discrete velocity integration:
\begin{equation}
  \hat{\bm{F}}_{\mathrm{b}}^{\,m=0}
  = \sum_{k} \bm{v}_k \cdot \bm{n}_{\mathrm{b}} \,\phi_{\mathrm{b}, k}
      \frac{\partial g}{\partial W_i}\, \Delta \Xi ,
  \label{eq:Fb_init}
\end{equation}
where $\phi_{\mathrm{b}}$ denotes the adjoint distribution at the boundary, computed by the upwind discretization of the kinetic adjoint equation~\eqref{eq:upwind_interface}. This provides a kinetically consistent macroscopic boundary flux that is used as the starting point for the adjoint NS solver.

Second, during the iterations of the adjoint NS solver, this boundary flux is further corrected by adding a linearized flux increment inferred from the kinetic boundary moment formulas. Denoting by $\hat{\bm{F}}_{\mathrm{b}}^{\,m}$ the boundary flux at iteration $m$, the update is written as
\begin{equation}
  \hat{\bm{F}}_{\mathrm{b}}^{\,m+1}
  \;\approx\;
    \hat{\bm{F}}_{\mathrm{b}}^{\,m}
    + \Delta \hat{\bm{F}}_{\mathrm{b}}^{\,m}
        \!\left(\Delta \hat{\bm{W}}\right),
  \label{eq:boundary_flux_update}
\end{equation}
where $\Delta \hat{\bm{F}}_{\mathrm{b}}^{\,m}(\Delta \hat{\bm{W}})$ denotes the linearized boundary flux increment, expressed as a function of the increment $\Delta \hat{\bm{W}}$ of the macroscopic adjoint variables at the boundary face. 
The explicit forms of the flux increments $\Delta \hat{\bm{F}}_{\mathrm{b}}^{\,m}$ for solid wall and Dirichlet boundaries are given in \ref{appendix_B_ajoint}, where the macroscopic increment $\Delta \hat{\bm{W}}$ at the boundary is obtained by interpolating the increments from the centers of cells adjacent to the boundary. 

By applying this predictor treatment to the adjoint NS boundary flux, the convergence speed of the adjoint GSIS can be significantly accelerated, as will be shown in Section~\ref{subsec:fast_convergence_adjoint_GSIS}.

\subsection{Sensitivity with respect to the boundary shape}\label{sec:sens_nod}

Once the primal flow variable $f$ and the adjoint variable $\varphi_{\rm w}$ have been obtained, the sensitivity of the objective functional $J$ with respect to the coordinates of the mesh nodes defining the solid boundary $\Gamma_{\rm w}$ is computed via Eq.~\eqref{eqn:sens_shape}, since $\mathcal L \equiv J$. The evaluation follows a strategy analogous to the discrete adjoint approach. The detailed derivation has been established in Ref.~\cite{yuan2025adjoint}, we only present the final computational formula here. 

First, the discretized forms of $J$ and $B_{\rm w}$ involved in Eq.~\eqref{eqn:sens_shape} are expressed as
\begin{equation}\label{eqn:numer_J}
\begin{aligned}
J& = \sum\limits_k {\sum\limits_l {{J_{l,k}}\Delta {\Xi _k}} }, \quad\text{wtih}\quad
{J_{l,k}} = {m_{J,k}}({{\bm v}_k} \cdot {{\bm n}_l}){f_{l,k}}{A_l}, \\
{B_{\rm{w}}}& = \sum\limits_{{{\vec v}_k} \in {\Xi ^ - }} {\sum\limits_l {{B_{{\rm{w}},l,k}}\Delta {\Xi _k}} },
\quad \text{with} \quad
{B_{{\rm{w}},l,k}} = {\varphi _{{\rm{w}},l,k}}  \left( {{{ f}_{l,k}} - {{g}_{{\rm w},l,k}}} \right){A_l},
\end{aligned}
\end{equation}
where the solid boundary $\Gamma _{\rm w}$ has been discretized into surface elements, indexed by $l$. For each element, $A_l,\bm x_l$ and $\bm n_l$ denote its area, centroid, and outward-pointing unit normal vector (pointing into the solid), respectively. All flow variables ${f}_{l,k}, {g}_{{\rm w},l,k}$ and adjoint variables $\varphi _{{\rm{w}},l,k}$ are evaluated at the element centroid. Note that ${{g}_{{\rm w},l,k}}$ here is calculated by the discretized version of Eq.~\eqref{eqn:formula_bgk_fdw}:
\begin{equation}
{{g}_{{\rm{w}},l,k}} = 
-\frac{{{{g}_{\rm{M}}}(1,\bm 0,{T_{\rm{w}}})}}{{  \sum\limits_{{{\bm v}_k} \in {\Xi ^ - }} {({{\bm v}_k} \cdot {{\bm n}_l}){g_{{\rm{M}}}}(1,\bm 0,{T_{\rm{w}}})\Delta {\Xi _k}} }}
\sum\limits_{{{\bm v}_k} \in {\Xi ^ + }} {({{\bm v}_k} \cdot {{\bm n}_l}){f_{l,k}}\Delta {\Xi _k}}.
\end{equation}

Then, the partial derivatives of  $J_{l,k}$ and $B_{{\rm{w}},l,k}$ with respect to the surface elements' attributes $A_l,\bm x_l$ and $\bm n_l$ are calculated as follows:
\begin{equation}\label{eqn:sens_J_itf}
\left.
\begin{aligned}
\frac{{\partial {J_{l,k}}}}{{\partial A_l}} &= {m_{J,k}} (\bm v_k \cdot \bm n_l){f_{l,k}},\\
\frac{{\partial {J_{l,k}}}}{{\partial {{\bm x}_l}}} &= {m_{J,k}} (\bm v_k \cdot \bm n_l){A_l}\frac{{\partial {f_{l,k}}}}{{\partial {{\bm x}_l}}},\\
\frac{{\partial {J_{l,k}}}}{{\partial {{\bm n}_l}}} &= {m_{J,k}}\bm v_k{A_l}{f_{l,k}},
\end{aligned}
\right\}
\end{equation}
and
\begin{equation}\label{eqn:sens_B_itf}
\left.
\begin{aligned}
\frac{{\partial {B_{{\rm{w}},l,k}}}}{{\partial {A_l}}} &= 0,\\
\frac{{\partial {B_{{\rm{w}},l,k}}}}{{\partial {{\bm x}_l}}} &= {{\varphi }_{{\rm{w}},l,k}}  \frac{{\partial {{f}_{l,k}}}}{{\partial {{\bm x}_l}}}{A_l} - {{ \varphi }_{{\rm{w}},l,k}}  {{\bar { g}}_{{\rm{w}},l,k}}{A_l}\left( {\sum\limits_{{{\bm v}_k} \in {\Xi ^ + }} {({{\bm v}_k} \cdot {{\bm n}_l})\frac{{\partial {f_{l,k}}}}{{\partial {{\bm x}_l}}}\Delta {\Xi _k}} } \right),\\
\frac{{\partial {B_{{\rm{w}},l,k}}}}{{\partial {{\bm n}_l}}} &=  - {{ \varphi }_{{\rm{w}},l,k}}  {{\bar { g}}_{{\rm{w}},l,k}}\left( {\sum\limits_{k } {{{\bm v}_k}{f_{l,k}}\Delta {\Xi _k}} } \right),
\end{aligned}
\right\}
\end{equation}
where the spatial derivative ${\partial {f_{l,k}}}/{{\partial {{\bm x}_l}}}$ will be calculated by the reconstruction using the Gauss formula.

With Eqs.~\eqref{eqn:sens_J_itf} and \eqref{eqn:sens_B_itf}, the total sensitivities of $J$ and $B_{\rm w}$ are obtained by summation over all velocity and surface elements according to Eq.~\eqref{eqn:numer_J}. The sensitivity of the Lagrangian $\cal L$ with respect to the surface elements' attributes $A_l,\bm x_l,\bm n_l$ then follows directly from Eq.~\eqref{eqn:sens_shape}.

Finally, after some geometric transformations, the above sensitivity can be further transformed into that with respect to the coordinates of the boundary mesh nodes. The transformation derivatives can be formulated as (assuming that the boundary surface elements are all triangular):
\begin{equation}
    \begin{aligned}
        \frac{\partial \bm x_{\rm c}}{\partial \bm r_m} &= \frac{1}{3}, \\
        \frac{{\partial A}}{{\partial {{\bm r}_m}}} &= {\rm{sign}}\left[\bm n \cdot ({{\bf v}_1} \times {{\bf v}_2}) \right] \frac{\bm n \times {{\bf v}_m}}{2},\\
        \frac{{\partial \bm n}}{{\partial {{\bm r}_m}}} &= {\rm{sign}}\left[\bm n \cdot ({{\bf v}_1} \times {{\bf v}_2}) \right]
        \frac{{\bf v}_m \times \bm n}{{2A}}\bm n.
    \end{aligned}
\end{equation}
In the above equation, $A,\bm x_{\rm c}$, and $\bm n$ are the area, centroid, and unit normal vector of a certain surface triangle element, respectively. $\bm r_m$ with $m=1,2,3$ is the coordinate of the triangle vertex. ${\bf v}_m$ is defined as \(\textbf{v}_1 = \bm{r}_3 - \bm{r}_2\), \(\textbf{v}_2 = \bm{r}_1 - \bm{r}_3\), and \(\textbf{v}_3 = \bm{r}_2 - \bm{r}_1\).
Then the sensitivity with respect to the coordinates of a certain solid boundary point can be calculated as:
\begin{equation}
\frac{{\partial {\cal L}}}{{\partial {{\bm r}_s}}} = \sum\limits_{l \in N(s)} {\left( {\frac{{\partial {\cal L}}}{{\partial {A_l}}}\frac{{\partial {A_l}}}{{\partial {{\bm r}_s}}} + \frac{{\partial {\cal L}}}{{\partial {{\bm x}_l}}}\frac{{\partial {{\bm x}_l}}}{{\partial {{\bm r}_s}}} + \frac{{\partial {\cal L}}}{{\partial {{\bm n}_l}}}\frac{{\partial {{\bm n}_l}}}{{\partial {{\bm r}_s}}}} \right)} ,
\end{equation}
where $s$ is the point index among all solid boundary points and $N(s)$ denotes its neighboring boundary surface elements.


\section{Boundary shape parameterization and mesh deformation}\label{sec:meshdeform}

As investigated in Ref.~\cite{yuan2025adjoint}, the sensitivity with respect to the boundary mesh nodes will suffer serious oscillations due to the discretization of the molecular velocity, and therefore some parameterizations should be performed to restrict the design variables to a design space of smooth, regular boundary shapes.
In this study, the parameterization of free form deformation (FFD) \cite{sederberg1986free} and spring smoothing \cite{batina1990unsteady, matsushima2002unstructured} are employed to deform the mesh while preserving its topological structure. FFD controls the geometric deformation, while spring smoothing ensures that the surrounding mesh adapts consistently to geometric changes.

\subsection{Free form deformation} \label{subsec:FFD}

The FFD method offers strong deformation capability, does not require the parameterization to conform to the initial shape, and preserves the continuity and smoothness of the original geometry, while also being user-friendly. As a result, it is widely used as a geometric parameterization technique in aerodynamic shape design for aircraft~\cite{duvigneau2006multi, palacios2012adjoint, economon2016su2}.

In its application, the FFD method embeds the object of study (i.e., the geometric shape to be deformed) into an control lattice and achieves geometric deformation by moving the positions of the control points. Therefore, in the process of parameterizing geometric shapes, the design parameters are the coordinates of the FFD control points.

First, we introduce an control lattice enclosing the target geometric shape to be deformed. The coordinates of its nodes, denoted as $\bm{x}$, can be expressed as
\begin{equation}
    \bm{x} (u, v, w) = \sum_{i=0}^{N_u-1} \sum_{j=0}^{N_v-1} \sum_{k=0}^{N_w-1} B_{i,d_u}(u) B_{j,d_v}(v) B_{k,d_w}(w) \bm{Q_{ijk}}.
    \label{eq:trivariate_b_spline}
\end{equation}
Here, $d_u$, $d_v$, and $d_w$ are the degrees of the basis functions, $\bm{Q}$ represents the coordinates of uniformly discretized control points within the lattice, $N_u$, $N_v$, and $N_w$ denote the number of control points along three directions, $u$, $v$, and $w$ represent the parameter space coordinates corresponding to the actual node coordinates. During the control lattice deformation process, the parameter space coordinates $(u_{ s},v_{ s},w_{ s})$ of the solid boundary points remain unchanged and can be determined by solving the least-squares problem $\min f = |\bm{x} - \bm{x_0}|^2$ at the initial time step. $B$ refers to the B-spline basis functions, generated by the following recursive rule (it is stipulated that any term on the right-hand side of the equation with a zero denominator is defined to be zero):
\begin{equation}
    B_{i,d}(u) = \frac{u - u_i}{u_{i+d} - u_i} B_{i,d-1}(u) + \frac{u_{i+d + 1} - u}{u_{i+d + 1} - u_{i+1}} B_{i+1,d-1}(u),
    \label{eq:b_spline_basis}
\end{equation}
with the initial conditions
\begin{equation}
    B_{i,0}(u) = 
    \begin{cases} 
        1, & u_i \leq u < u_{i+1}, \\ 
        0, & \text{otherwise},
    \end{cases}
\end{equation}
where \(u_i\) is a knot in the open uniform knot vector normalized to \([0, 1]\). 

Furthermore, we need to transform the derivatives with respect to the vertices into derivatives concerning the coordinates of FFD control points \(\bm{Q}\):
\begin{equation}
    \frac{\partial \bm r}{\partial \bm{Q}_{ijk}} = B_{i,d_u}(u_{ s}) B_{j,d_v}(v_{ s}) B_{k,d_w}(w_{ s}).
\end{equation}
Finally, the sensitivity of the Lagrangian \(\mathcal L\) with respect to the coordinates of FFD control points can be computed as:
\begin{equation}
\frac{{\partial {\cal L}}}{{\partial {{\bm Q}_{ijk}}}} = \sum\limits_s {\frac{{\partial {\cal L}}}{{\partial {{\bm r}_s}}}\frac{{\partial {{\bm r}_s}}}{{\partial{{\bm Q}_{ijk}}}}}.
\end{equation}
where $s$ denotes the index of all nodes on the solid boundaries to be designed.

\subsection{Spring smoothing} \label{subsec:Spring_Smoothing}

As a method of physical structure analogy, the spring smoothing analogy assumes that mesh nodes are connected by springs. The entire computational mesh is regarded as a structural system composed of multiple springs. The forces in the springs and the displacements of the nodes follow Hooke's Law~\cite{batina1990unsteady}.
The resultant force $\boldsymbol{F}_{i}$ on node $i$ can be expressed as:
\begin{equation}
    \boldsymbol{F}_{i} = \sum_{j \in N_{i}} k_{ij} \left( \Delta\boldsymbol{x}_{j} - \Delta\boldsymbol{x}_{i} \right),
\end{equation}
where $N_i$ is the set of all nodes connected to node $i$ within the computational domain, $\Delta x_i$ and $\Delta x_j$ are the displacement vectors of nodes $i$ and $j$, respectively, and $k_{ij}$ is the spring stiffness coefficient between nodes $i$ and $j$.
After the mesh points are moved, the resultant force at the new position is zero, i.e., $\boldsymbol{F}_{i} = 0$. 

The successive over-relaxation with $\omega=1.8$ is employed for the iterative solution.
The displacement of node $i$ at the $m$-th iteration is given by:
\begin{equation}
    \Delta \boldsymbol{x}_{i}^{m} = \Delta \boldsymbol{x}_{i}^{m-1} + \omega \left(\frac{\sum_{j \in L(i)} k_{ij} \Delta \boldsymbol{x}_{j}^{m} + \sum_{j \in U(i)} k_{ij} \Delta \boldsymbol{x}_{j}^{m-1}}{\sum_{j \in N_{i}} k_{ij}} - \Delta \boldsymbol{x}_{i}^{m-1} \right),
\end{equation}
where $L(i)$ and $U(i)$ are subsets of the set $N_i$ of neighboring node indices of node $i$. Specifically, $L(i)$ contains the neighboring node indices less than $i$, while $U(i)$ contains those greater than $i$. 
Finally, the new position of node $i$ is obtained as:
\begin{equation}
    \boldsymbol{x}_{i}^{\text{new}} = \boldsymbol{x}_{i}^{\text{old}} + \Delta \boldsymbol{x}_{i}^{m}.
\end{equation}

To achieve more reasonable global mesh deformation, a wall-distance function was introduced when setting the spring stiffness coefficient~\cite{matsushima2002unstructured}:
\begin{equation}
    \label{eq:spring-kij}
    k_{ij} = {d^{-1}_{ij}} {l_{ij}^{-2}},
\end{equation}
where \(d_{ij}\) is the shortest distance from the midpoint of the line connecting nodes \(i\) and \(j\) to the deformable geometry (i.e., the wall), and \(l_{ij}\) is the length of the line connecting nodes \(i\) and \(j\).
This helps neighboring nodes that are very close to each other maintain their relative positions during the smoothing iteration process, thereby better preserving geometric features and details of local region.



\section{Numerical Results}\label{sec:num_example}

In this section, numerical simulations are conducted to evaluate the effectiveness of the proposed shape-optimization method in minimizing the drag of a 3D object in rarefied gas flows.

\begin{figure}[t]
    \centering
    \includegraphics[width=1\linewidth]{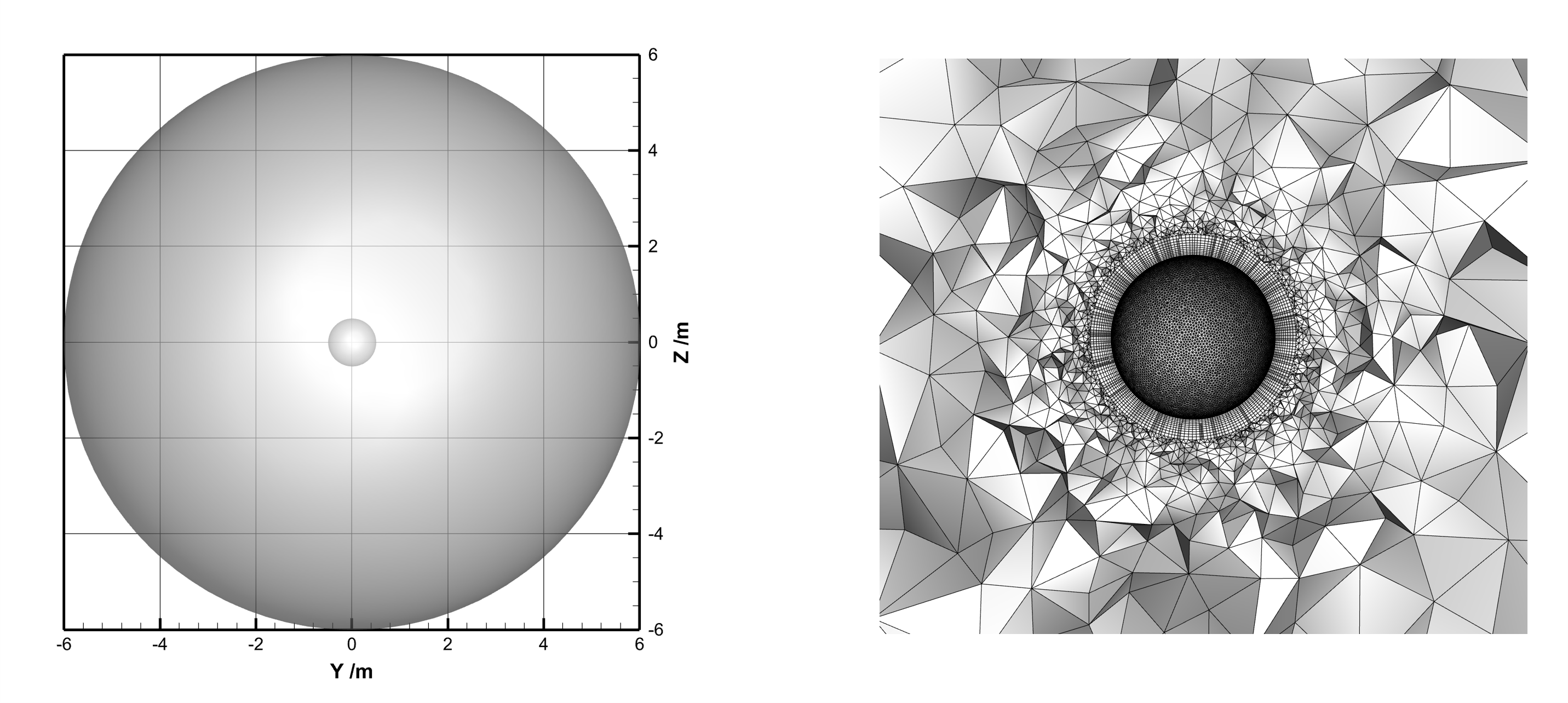}
    \caption{(Left) Schematic of flow simulation around a sphere with its center at the origin (0, 0, 0). 
    (Right) Cross-section of the hybrid computational mesh on the plane $x=0$, consisting of prismatic layers near the inner spherical wall and tetrahedral elements in the outer region.
    }
    \label{fig:Sphere_flow_and_mesh}
\end{figure}

\subsection{Validation of the sensitivity} \label{subsec:validation_adjoint_GSIS}

The sensitivity analysis of design variables with respect to boundary node coordinates has been detailed in Ref.~\cite{yuan2025adjoint}. In adjoint-based shape optimization, discretization of the molecular velocity space $\bm{v}$ often introduces oscillations in the sensitivities, which are difficult to eliminate. Parametrization methods effectively smooth these oscillations—acting much like a low-pass filter—and yield smoother optimization results. Here,  the sensitivity of the design variables with respect to the FFD control-point coordinates is examined.


The simulation domain consists of a large sphere with radius 6 and a small sphere with radius 0.5 (normalized by $ l_{\rm ref}$). The control lattice is defined as a cubic lattice of side length 1.6, centered at the origin and fully enclosing the small sphere. As shown in Eq.~\eqref{eq:trivariate_b_spline}, B-spline basis functions of degree 3 (i.e., \(d_u = d_v = d_w = 3\)) are employed, with 8 discrete points (\(N_u = N_v = N_w = 8\), which should exceed the degree of the basis functions), resulting in a total of 1,536 design variables.
The large sphere's surface is the far field \((f = g_m(\rho_\infty, \bm{u}_\infty, T_\infty))\), with an inlet velocity of \(u_z = 2\) Mach. The small sphere's surface is a solid wall with a diffuse reflection boundary condition \((T_w = T_\infty)\). 
The simulation domain is discretized using tetrahedral and triangular prism elements. For $\text{Kn}$ = 0.1 and 0.01, the computational domain comprises 224,864 and 407,263 cells, respectively, with mesh refinement at the wall. The first-layer grid height at the wall is set to 0.01 and 0.001, respectively, to satisfy the simulation accuracy requirements.

\begin{figure}[t]
    \centering
   {
    \label{fig:Sphere_Sensitivity_Kn0d1_Qz}
        {\includegraphics[width=0.45\linewidth]{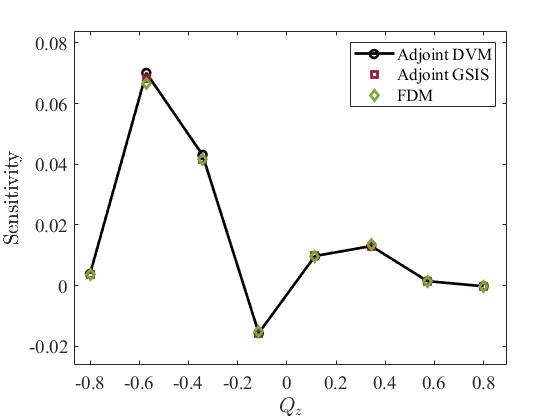}} }
    {
    \label{fig:Sphere_Sensitivity_Kn0d01_Qz}
        {\includegraphics[width=0.45\linewidth]{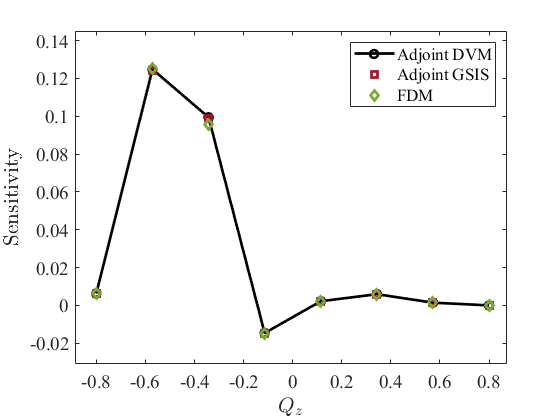}} }
    \caption{The sensitivity of the wall drag force to the $z$-coordinate of the FFD control points (\(Q_x = Q_y = -0.114286\)) in the numerical simulation of flow around a sphere at Ma = 2, with Kn=0.1 (left) and 0.01 (right). }
    \label{fig:Sphere_Sensitivity_Qz}
\end{figure}

The drag force acting on the solid wall serves as the objective function to verify the accuracy of the sensitivity concerning the FFD control point coordinates. The sensitivity computed via the GSIS is compared with that from the central finite-difference method:
\begin{equation}
    \left(\frac{\partial J}{\partial (Q_{ijk})_q}\right)^{\mathrm{FDM}} = \frac{J\left[(Q_{ijk})_q + \epsilon \right] - J\left[(Q_{ijk})_q - \epsilon \right]}{2\epsilon}, 
\end{equation}
where \(\epsilon\) is the perturbation amplitude, set to \(10^{-3}\) in this case. A uniform orthogonal discretization of \(24 \times 24 \times 30\) is applied to the molecular velocity space $\bm{v}$, truncated within \([-8v_0, 8v_0] \times [-8v_0, 8v_0] \times [-10v_0, 10v_0]\).



Two free-stream Knudsen numbers, $\text{Kn}=0.1$ and $\text{Kn}=0.01$, are tested for the initial geometry. The sensitivity of the wall drag force to the $z$-coordinate of the FFD control points ($Q_x = Q_y = -0.114286$) is analyzed. As shown in Fig.~\ref{fig:Sphere_Sensitivity_Qz}, the results obtained using the conventional DVM \eqref{eqn:implicit_kinetic0} of adjoint equations and the GSIS method exhibit a maximum relative error of no more than 4\% when compared to those from the central finite-difference method, validating the accuracy of the simulation program. 

\subsection{The fast convergence of GSIS for adjoint equation} \label{subsec:fast_convergence_adjoint_GSIS}


To compare the convergence rates of the conventional DVM and the proposed GSIS method, we consider the sensitivity of the wall drag force with respect to the free-stream density at the far-field boundary, defined as
\begin{equation}
    S_{\rho_{\infty}} = \sum\limits_{{{\vec v}_k} \in {\Xi ^ - }} {\sum\limits_{l \in \Gamma _{\rm{d}}} ({\bm{v}_k \cdot \bm{n}_l) \phi _{{\rm{d}}, l, k} g_{{\rm{d}}, l, k} A_l \Delta {\Xi _k}} },
\end{equation}
where $g_{\rm{d}} = g_{\text{M}}(1, \bm{u}_\infty, T_\infty)$. 
The adjoint variables evolve from the wall toward the far field, and $S_{\rho_{\infty}}$ effectively serves as an indicator of the convergence rate.

\begin{figure}[t]
    \centering
{\includegraphics[width=0.45\linewidth]{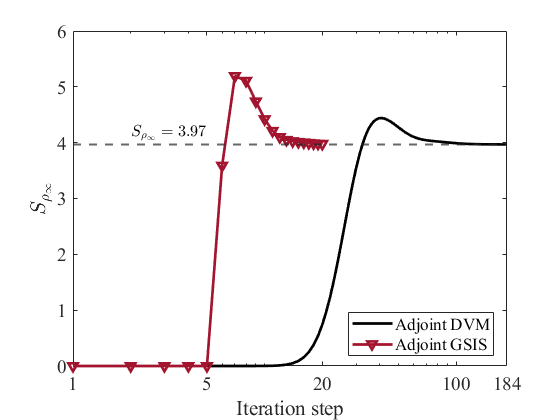}} 
{\includegraphics[width=0.45\linewidth]{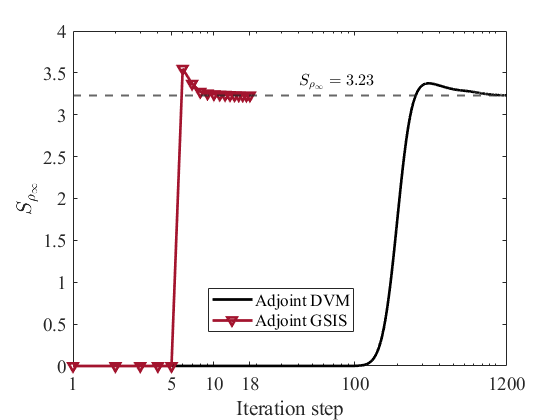}} 
    \caption{The variation of \( S_{\rho_{\infty}} \) with iteration steps for the adjoint equations using the conventional DVM and GSIS methods, with Kn=0.1 (left) and 0.01 (right). The first five steps in the GSIS method involve the initial field evolution using the conventional DVM.}
    \label{fig:Sphere_Ma2_rho_Sen_step_cmp}
\end{figure}


Fig.~\ref{fig:Sphere_Ma2_rho_Sen_step_cmp} shows the variation of $S_{\rho_{\infty}}$ with iteration steps when the adjoint equation is solved by the conventional DVM and GSIS. In both test cases, the GSIS algorithm converges within 20 steps. Notably, for Kn=0.01, GSIS achieves nearly a two-orders-of-magnitude reduction in iterations compared to conventional DVM. This underscores GSIS’s advantage in transition and slip flows, consistent with previous studies on the primal kinetic equation~\cite{su2020can,zhang2024efficient}. 
The key to its rapid convergence lies in the macroscopic synthetic equation, whose diffusion-type operator enables global information exchange and thus propagates the wall's influence to the far field even after a single application.

\subsection{The optimizing process and final shape}

After validating the accuracy of the adjoint-based sensitivities in Section~\ref{subsec:validation_adjoint_GSIS} and demonstrating the fast convergence of the adjoint GSIS solver in Section~\ref{subsec:fast_convergence_adjoint_GSIS}, we now apply the complete optimization framework to the 3D configuration introduced in Fig.~\ref{fig:Sphere_flow_and_mesh}. The design variables are the coordinates of the FFD control points defined in Section~\ref{subsec:FFD}, and the mesh is updated by the wall-distance-based spring smoothing in Section~\ref{subsec:Spring_Smoothing}, so that the topology of the hybrid prism-tetrahedral grid is preserved throughout the optimization. The objective functional is the drag force acting on the solid wall, while the volume constraint~\eqref{eq:volume-constraint} is enforced so that the optimized body remains comparable in size to the initial sphere.


\begin{figure}[t!]
    \centering
    {\includegraphics[width=0.45\linewidth]{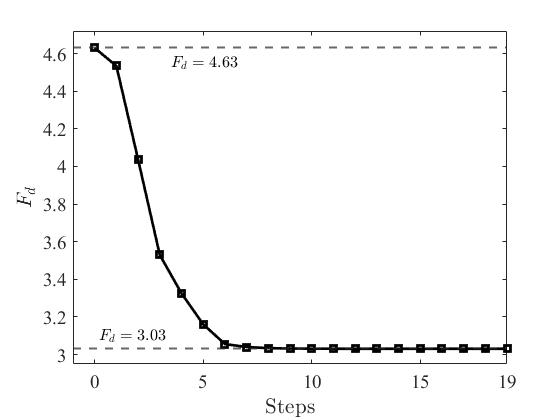}} 
    {\includegraphics[width=0.45\linewidth]{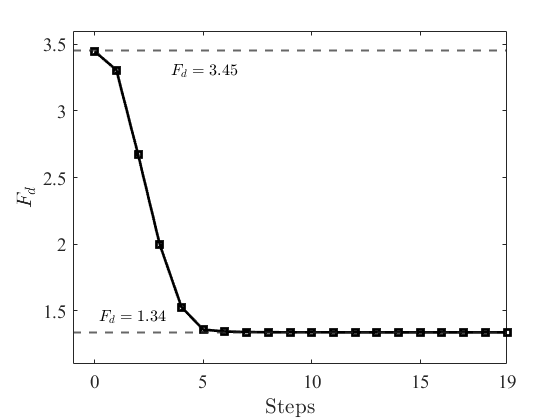}} 
    \caption{Evolution of sphere drag with optimization iterations under different Knudsen numbers at $\text{Ma} = 2$, achieving drag reductions of 34.5\% at Kn=0.1 (left) and 61.1\% at Kn=0.01 (right).}
    \label{fig:Sphere_Ma2_Op_Step}
\end{figure}

\begin{figure}[!t]
    \centering
    {\includegraphics[width=0.3\linewidth]{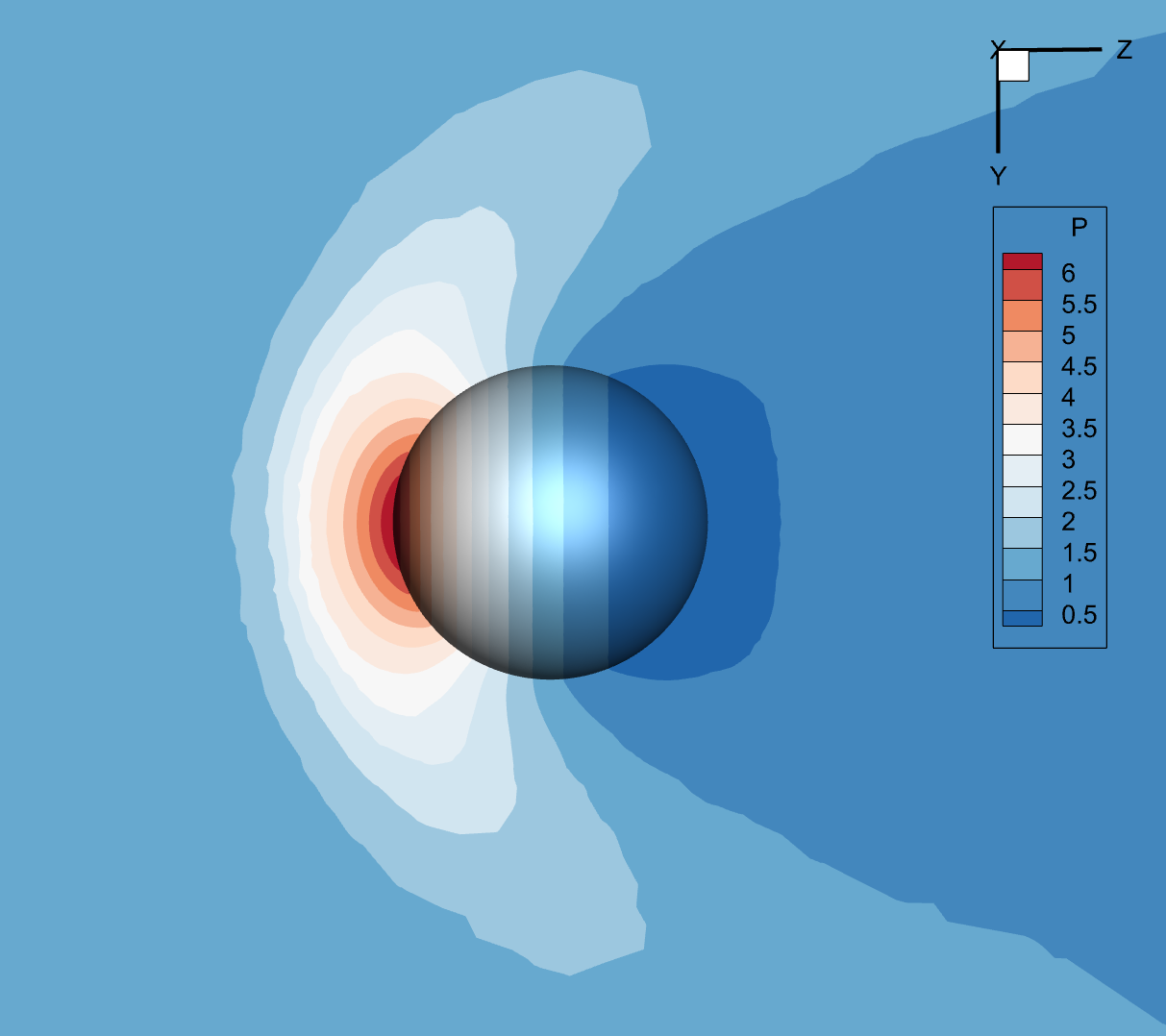}} 
    {\includegraphics[width=0.3\linewidth]{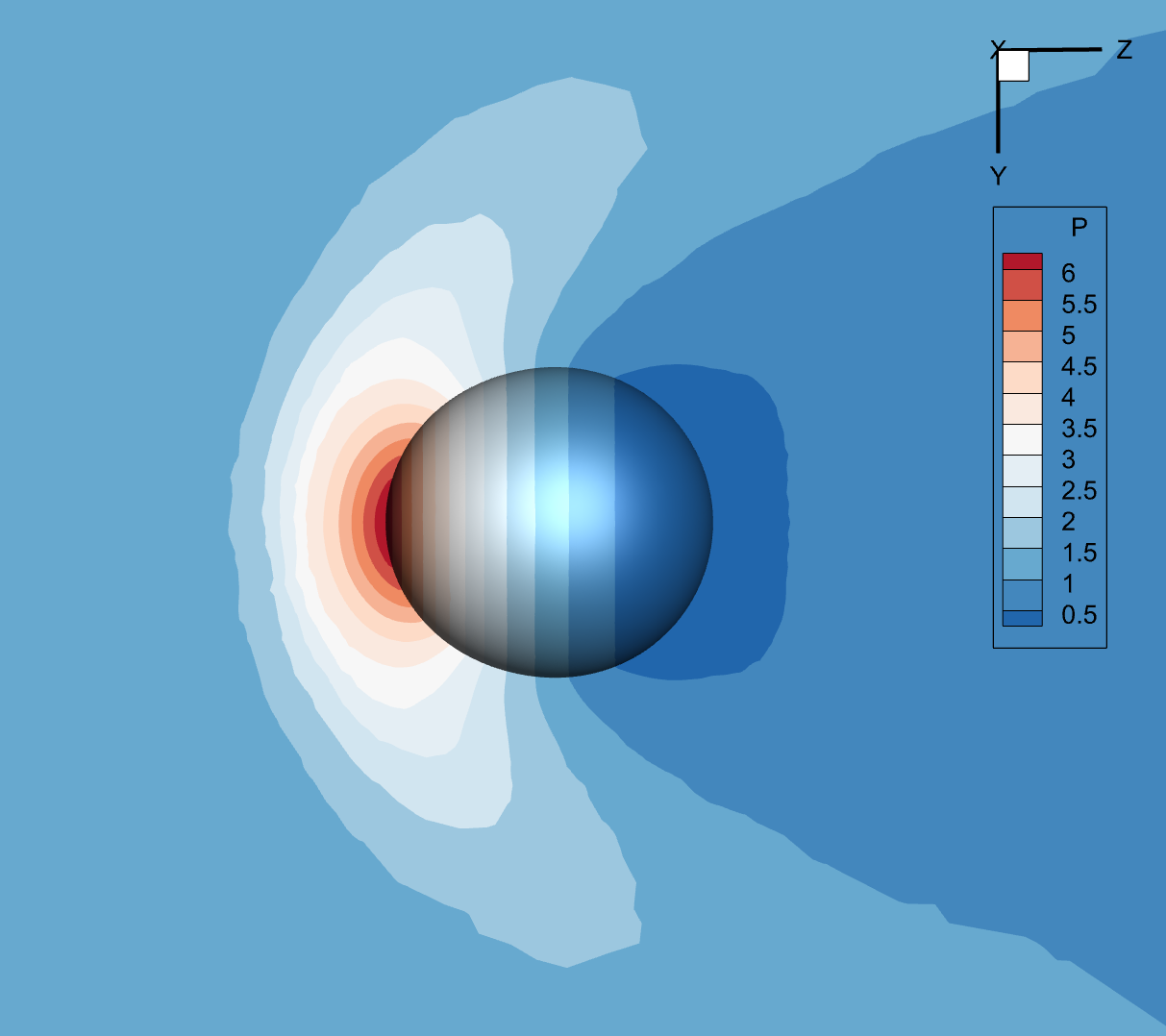}}
    {\includegraphics[width=0.3\linewidth]{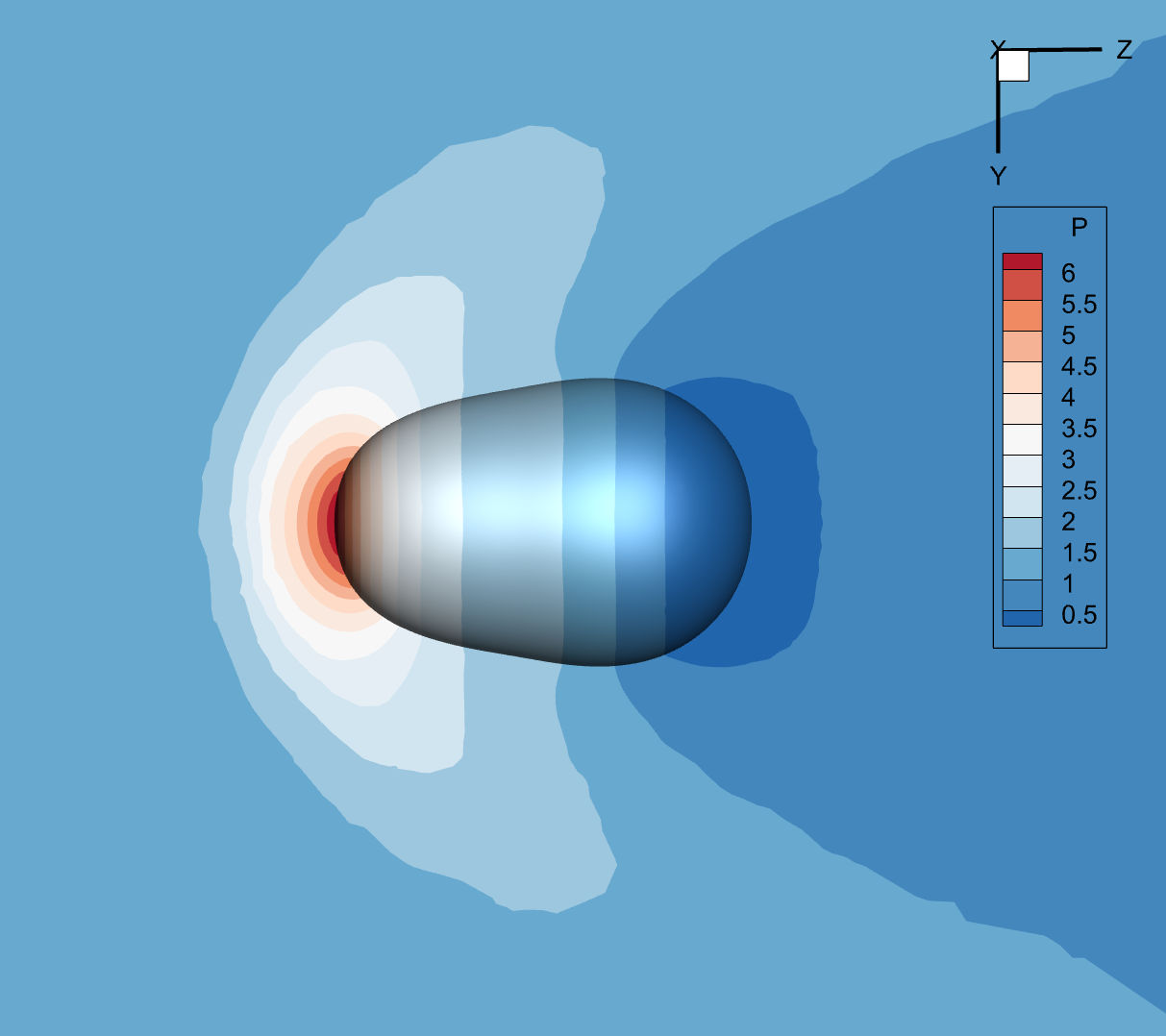}}\\
    \vspace{0.3cm}
    {\includegraphics[width=0.3\linewidth]{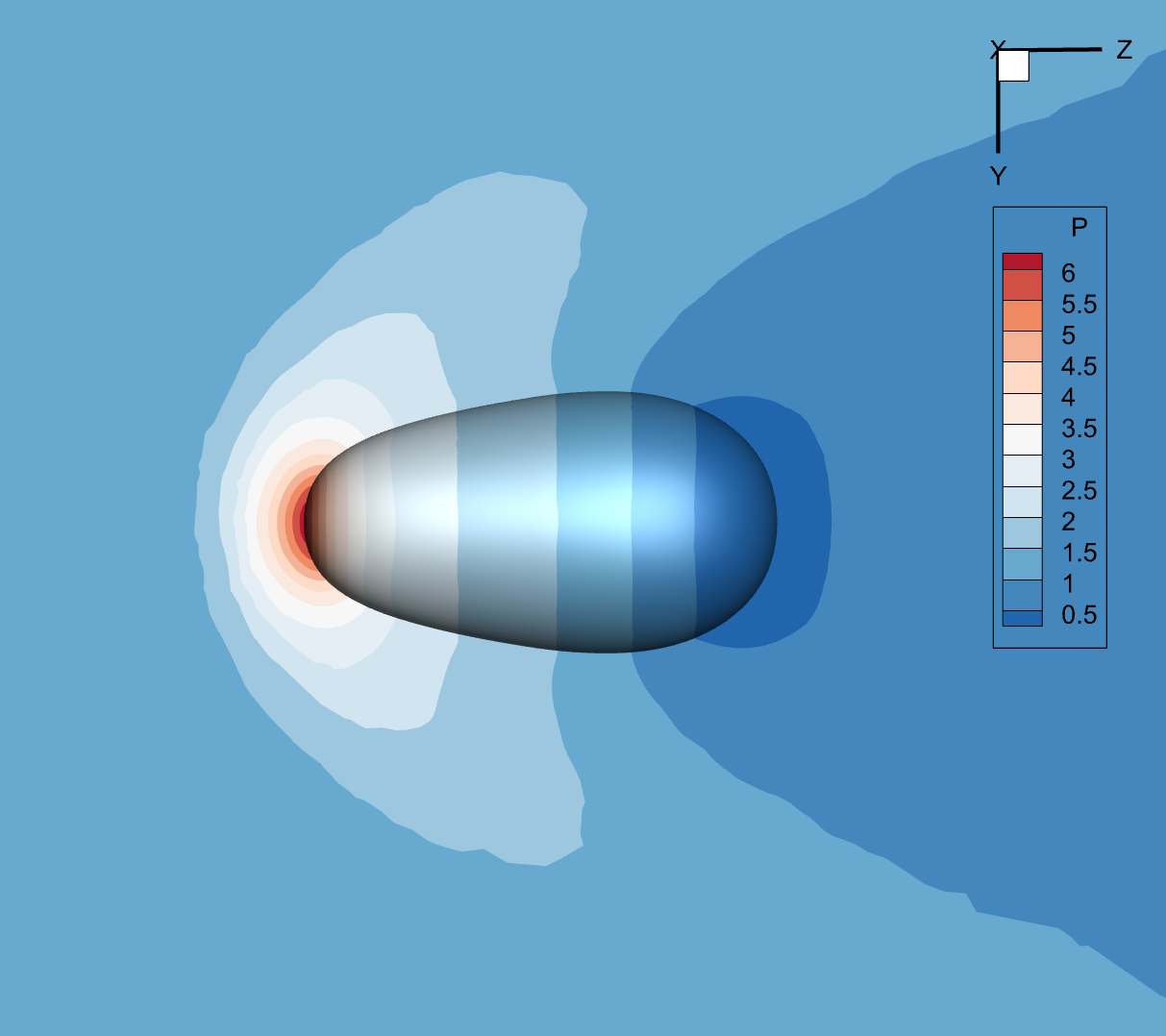}} 
    {\includegraphics[width=0.3\linewidth]{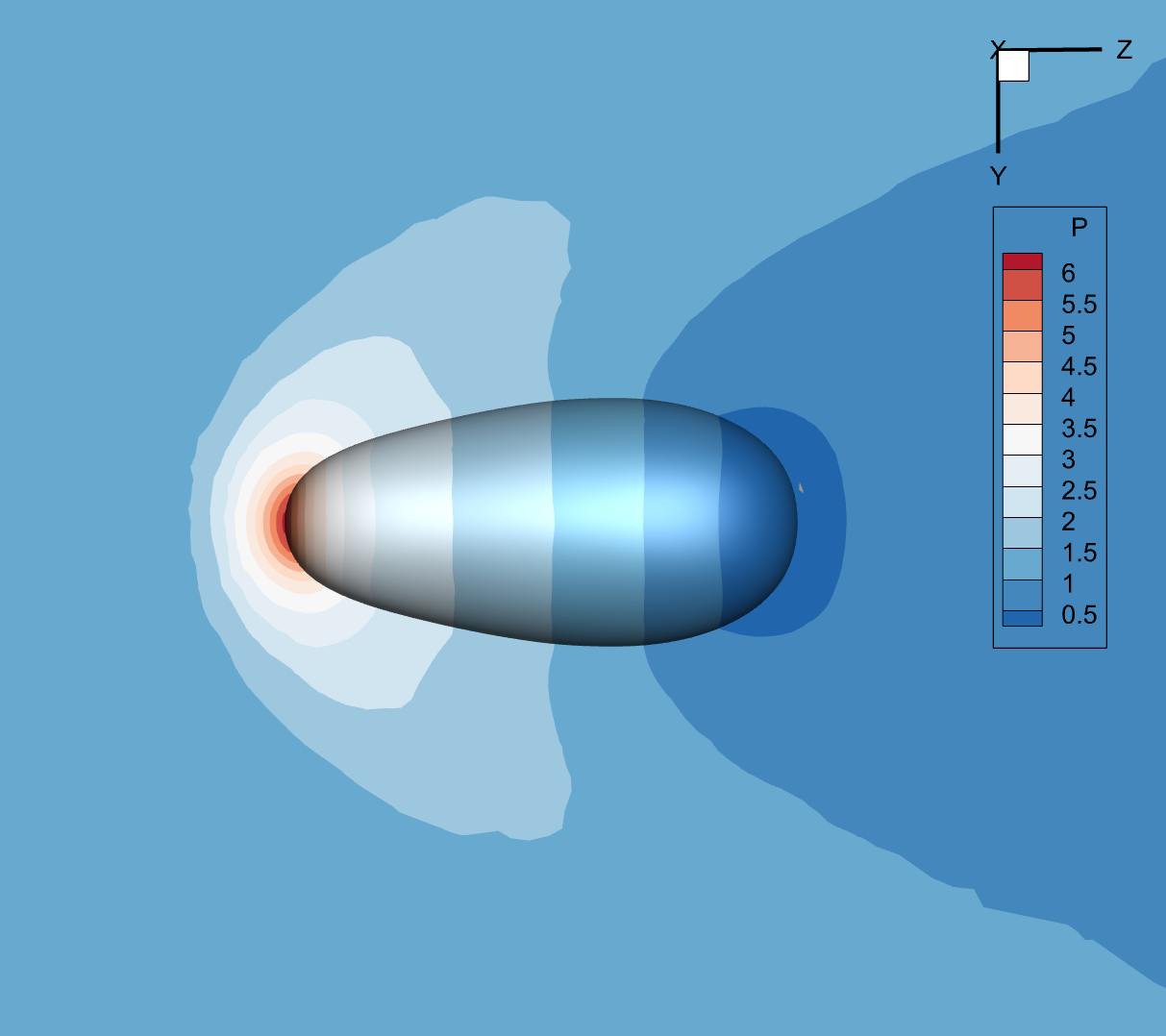}} 
    {\includegraphics[width=0.3\linewidth]{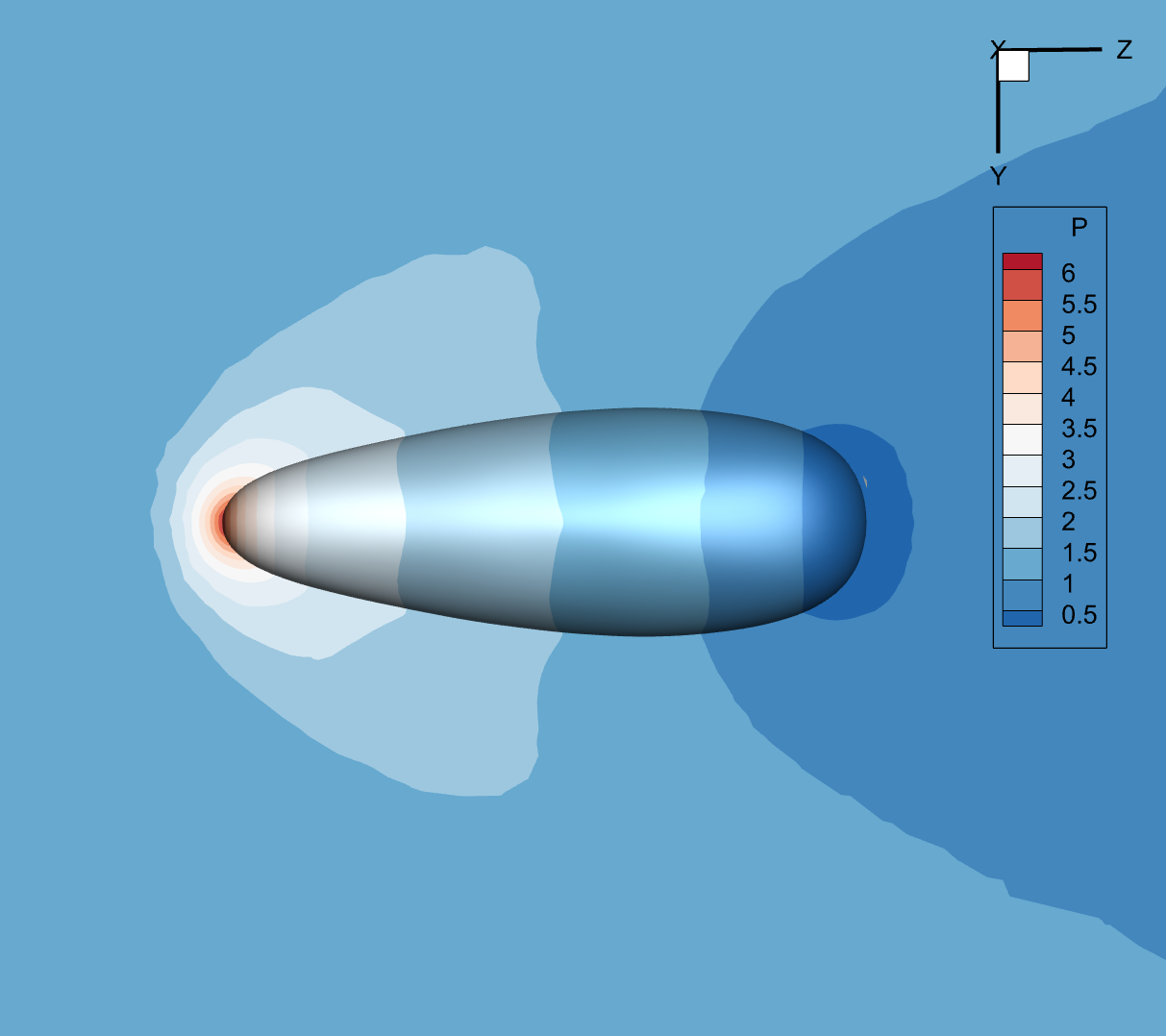}} \\
    \vspace{0.3cm}
    {\includegraphics[width=0.3\linewidth]{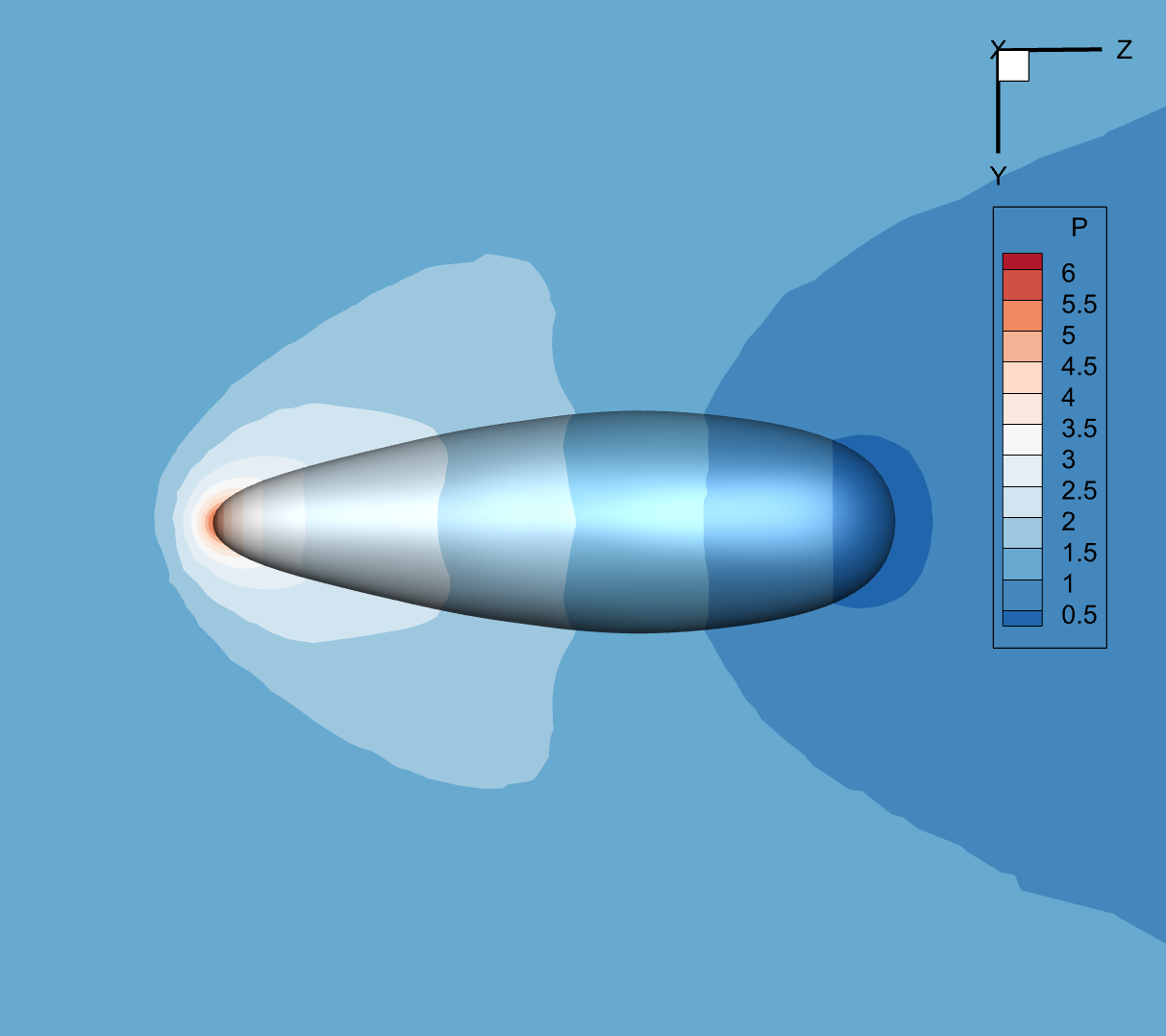}} 
    {\includegraphics[width=0.3\linewidth]{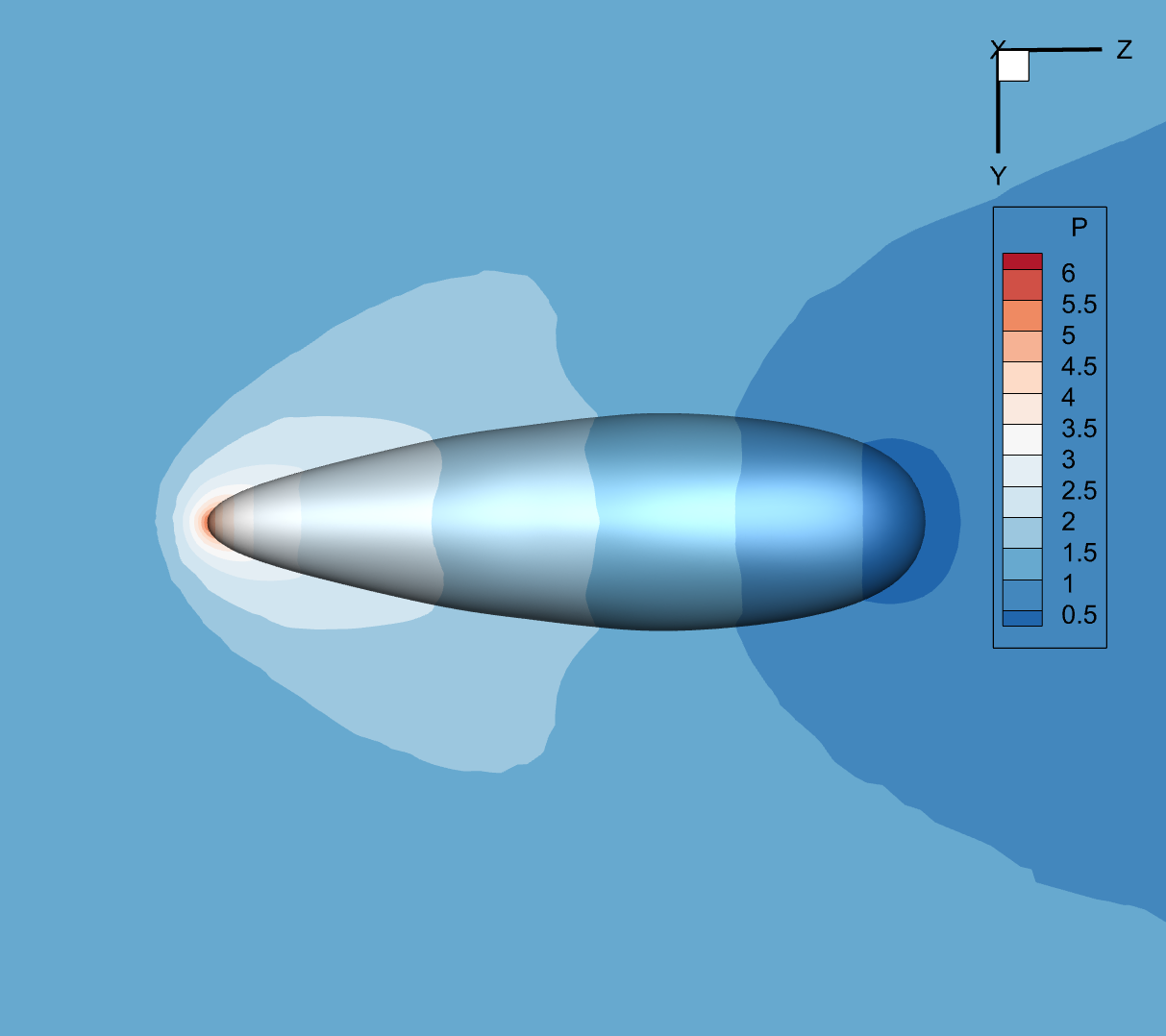}} 
    {\includegraphics[width=0.3\linewidth]{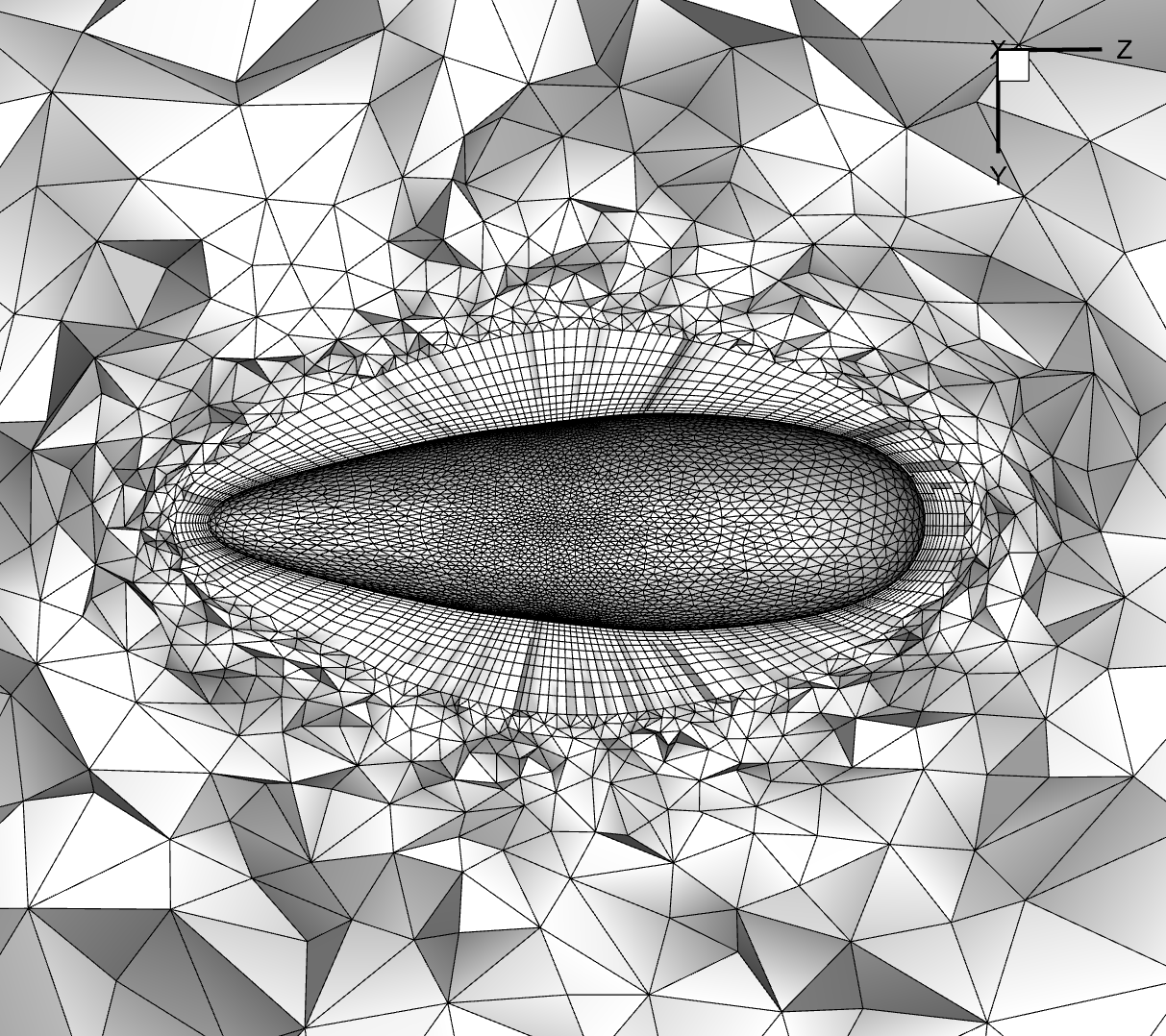}} 
    \caption{Evolution of the body shape and corresponding pressure field during optimization at $\mathrm{Kn}=0.1$ and $\mathrm{Ma}=2$. From left to right and from top to bottom, the panels show the initial shape, the shapes after the first to sixth optimization iterations, and the final optimized shape together with the deformation of the computational grid at the end of optimization in the $y$-$z$ plane with $x=0$.
    }
    \label{fig:Sphere_Ma2_Kn0d1_evolution}
\end{figure}

\begin{figure}[!t]
    \centering
    {\includegraphics[width=0.3\linewidth]{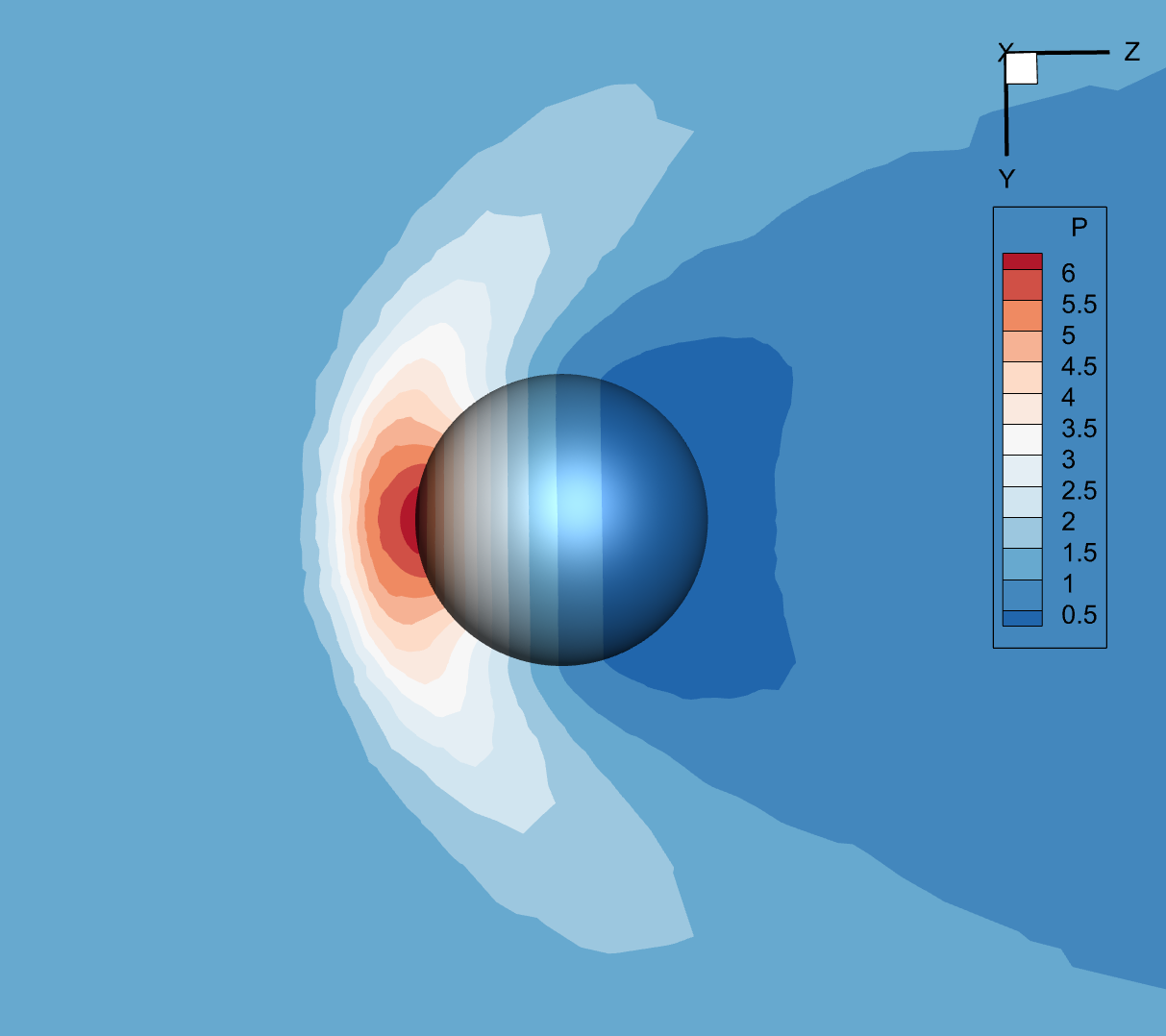}} 
    {\includegraphics[width=0.3\linewidth]{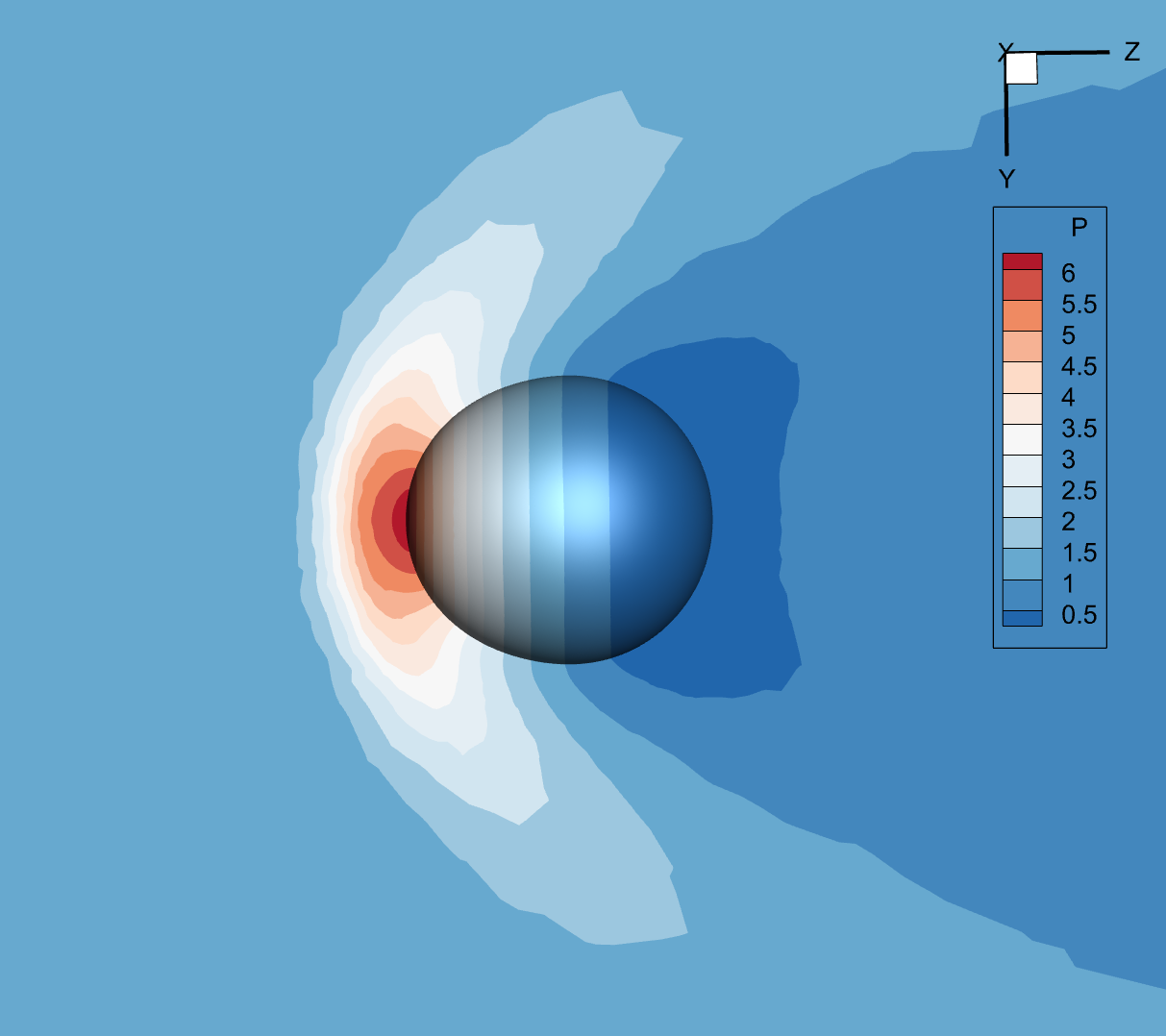}}
    {\includegraphics[width=0.3\linewidth]{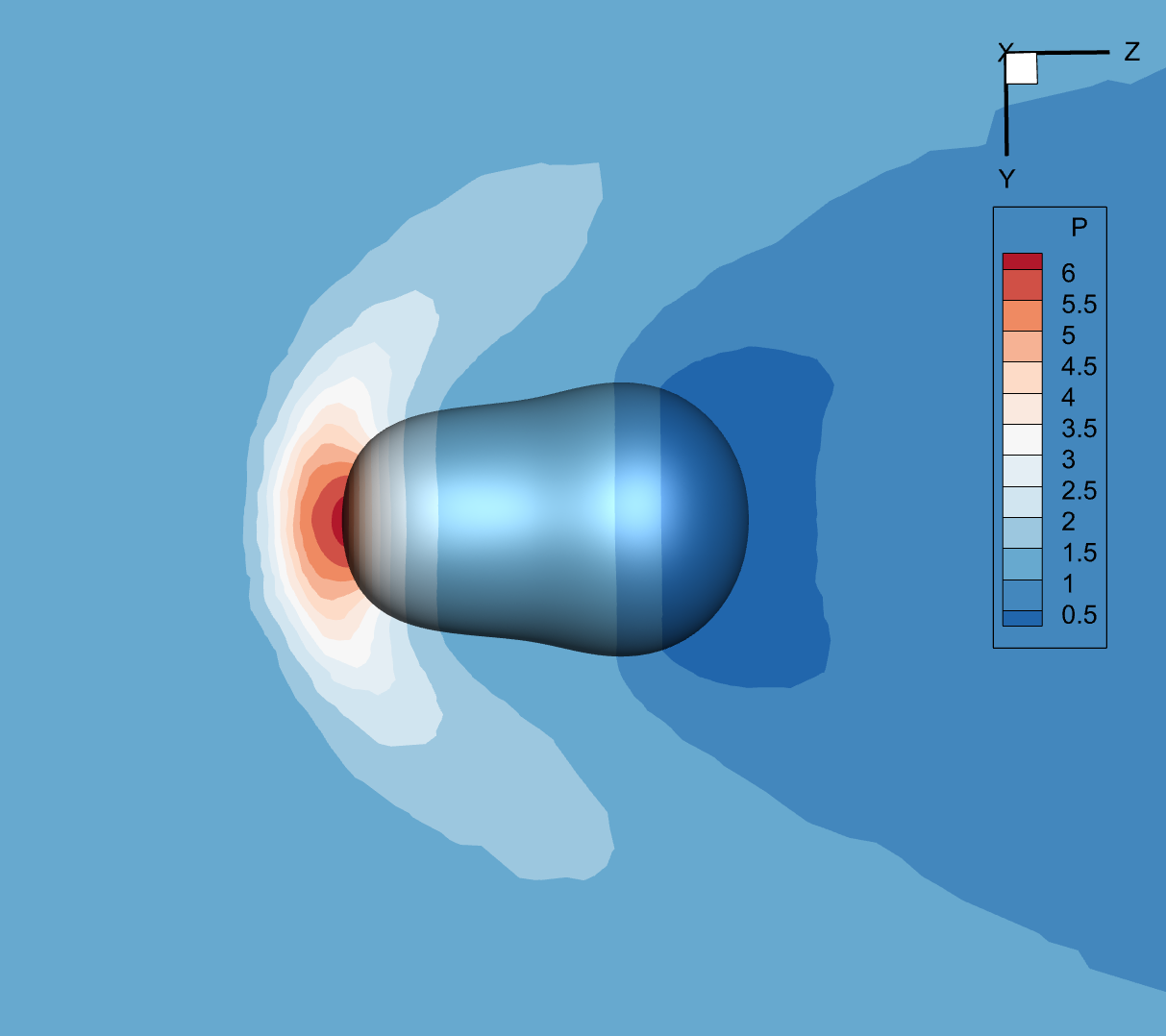}}\\
    \vspace{0.3cm}
    {\includegraphics[width=0.3\linewidth]{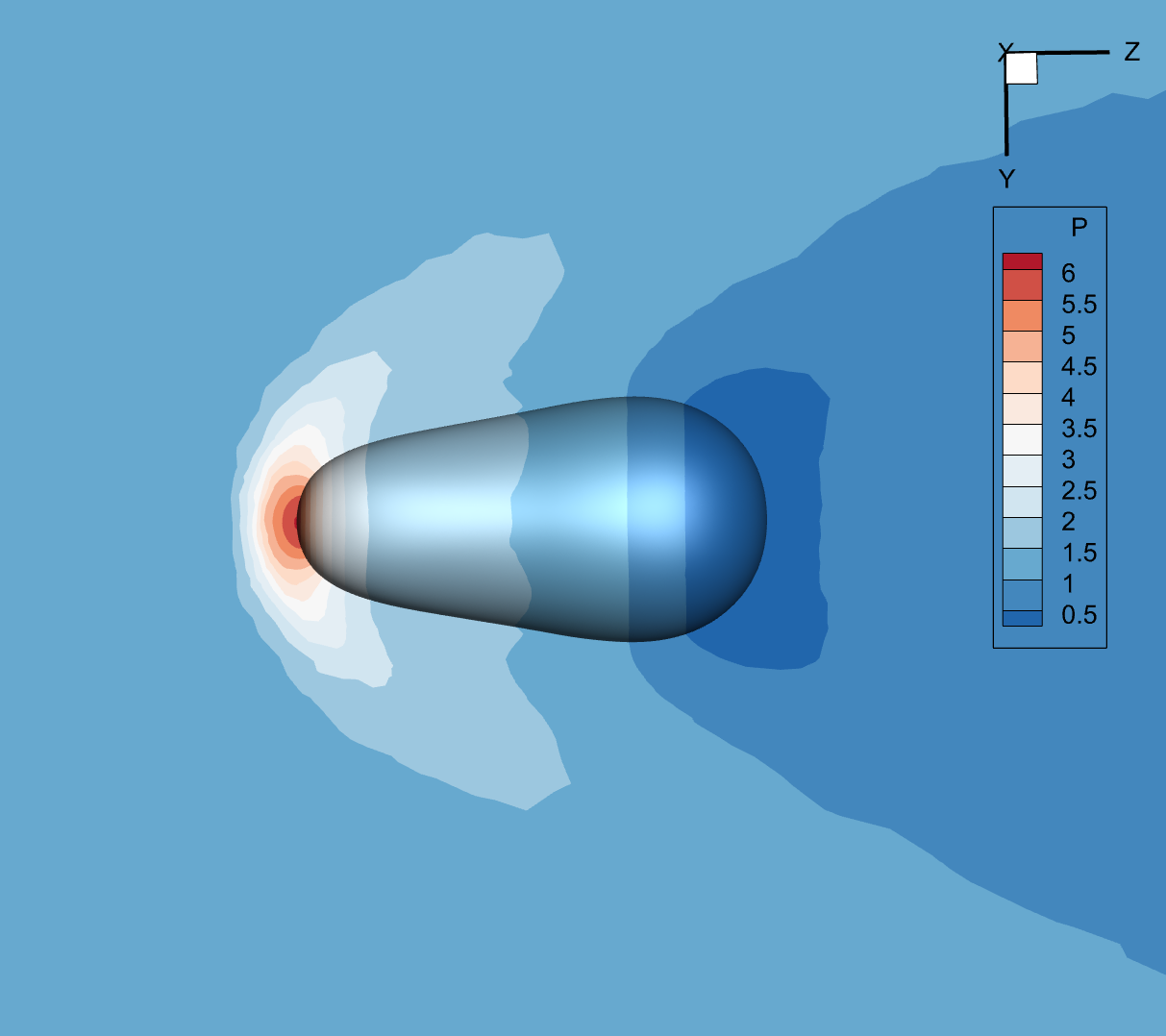}} 
    {\includegraphics[width=0.3\linewidth]{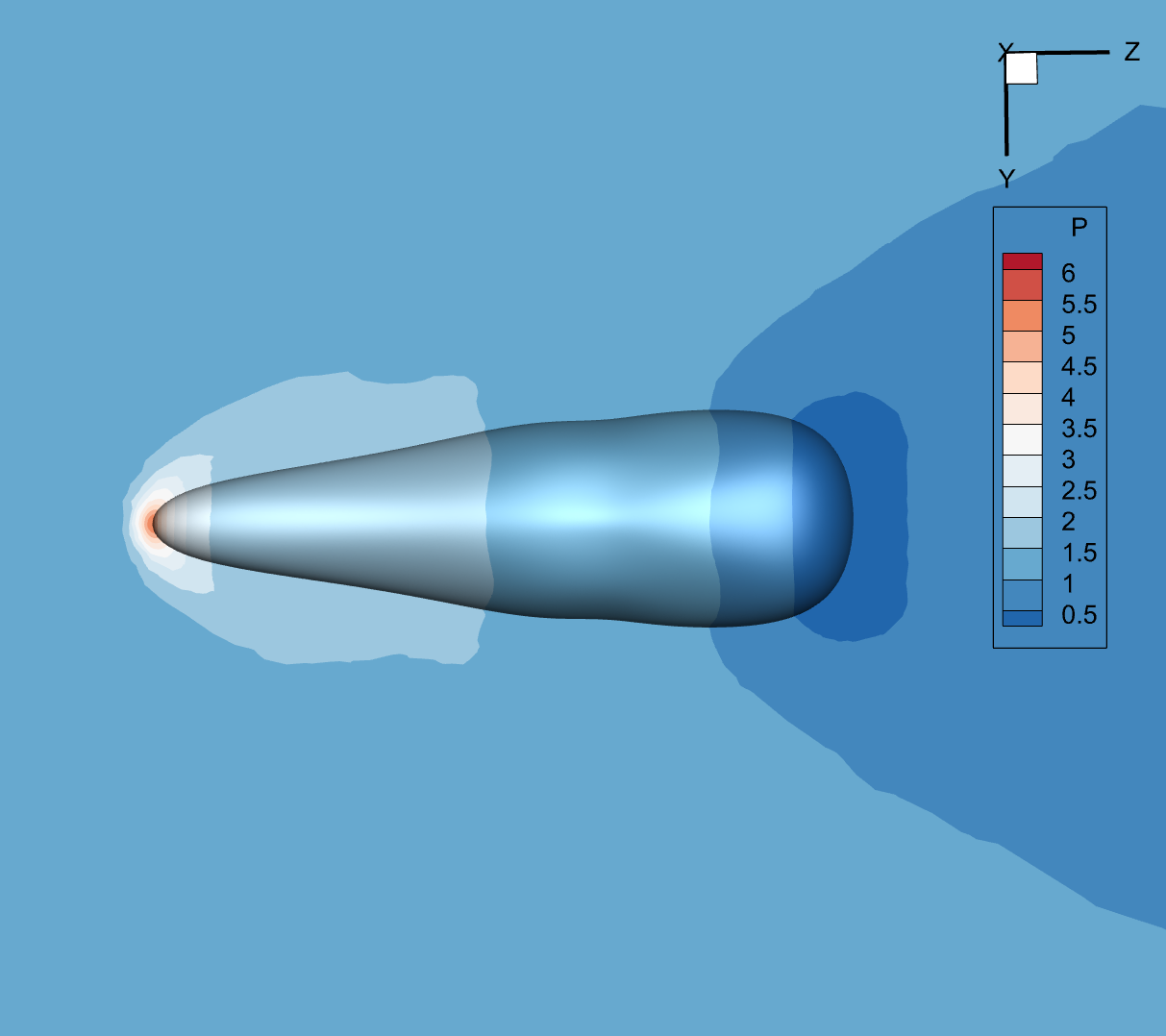}} 
    {\includegraphics[width=0.3\linewidth]{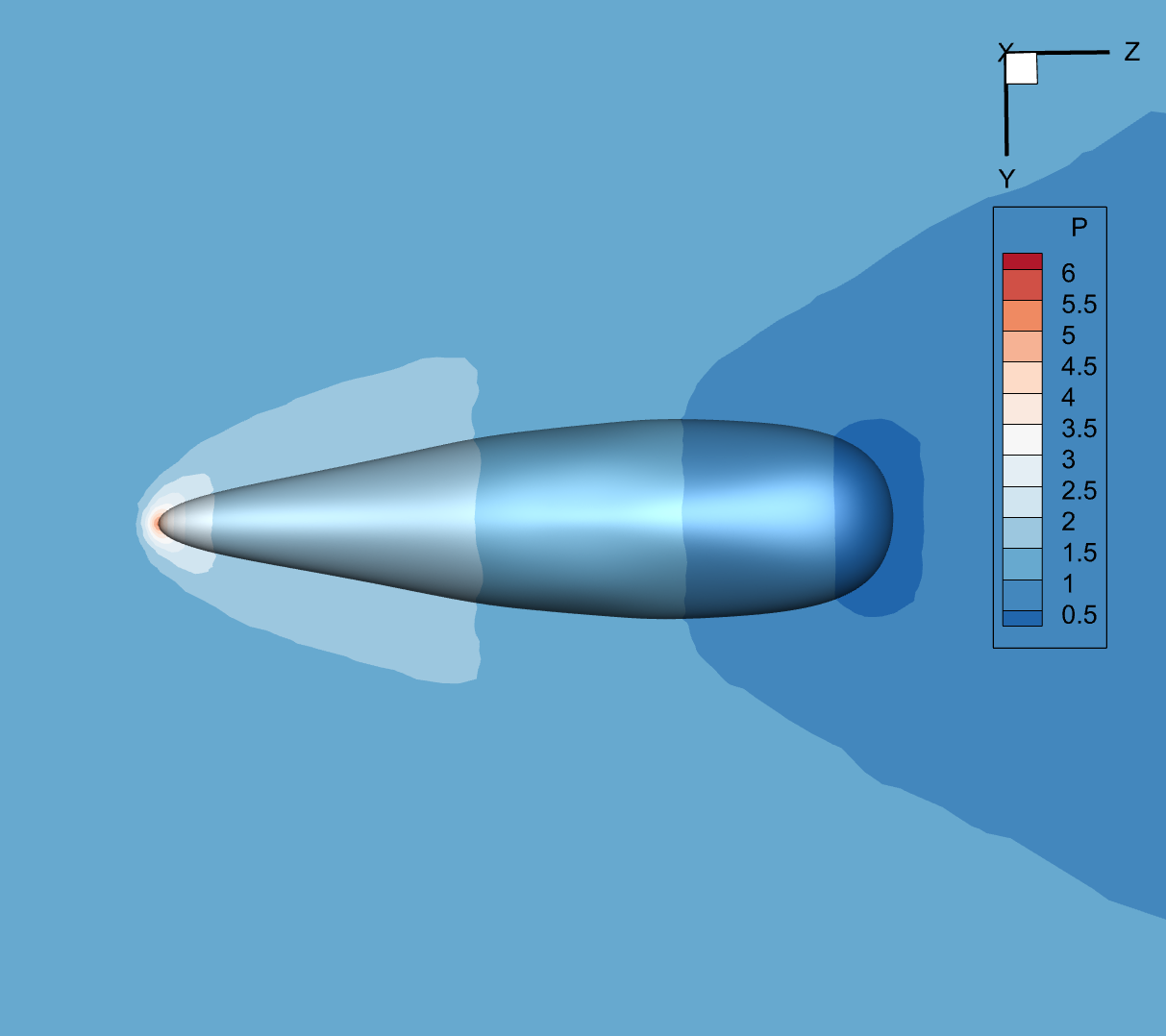}} \\
    \vspace{0.3cm}
    {\includegraphics[width=0.3\linewidth]{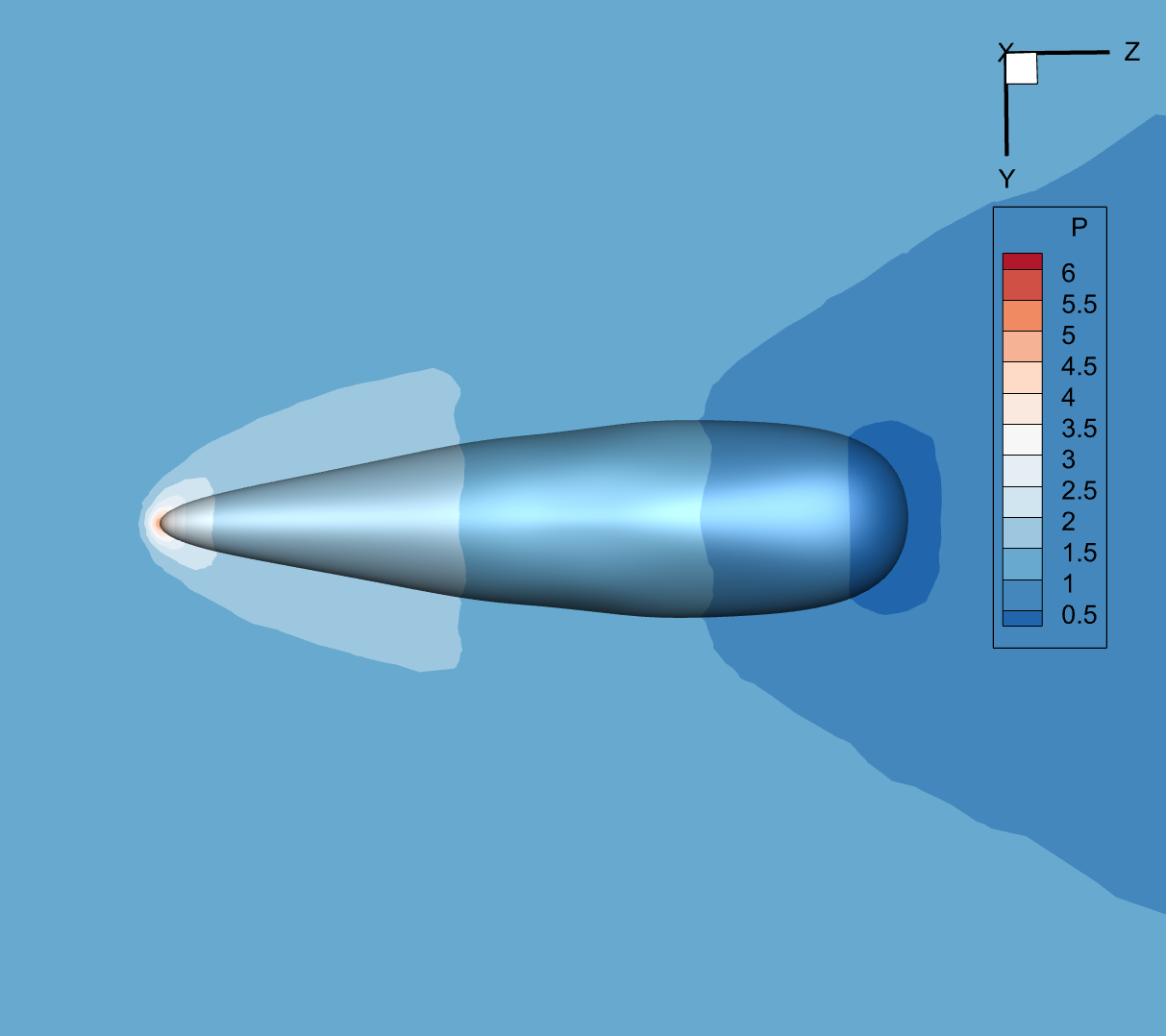}} 
    {\includegraphics[width=0.3\linewidth]{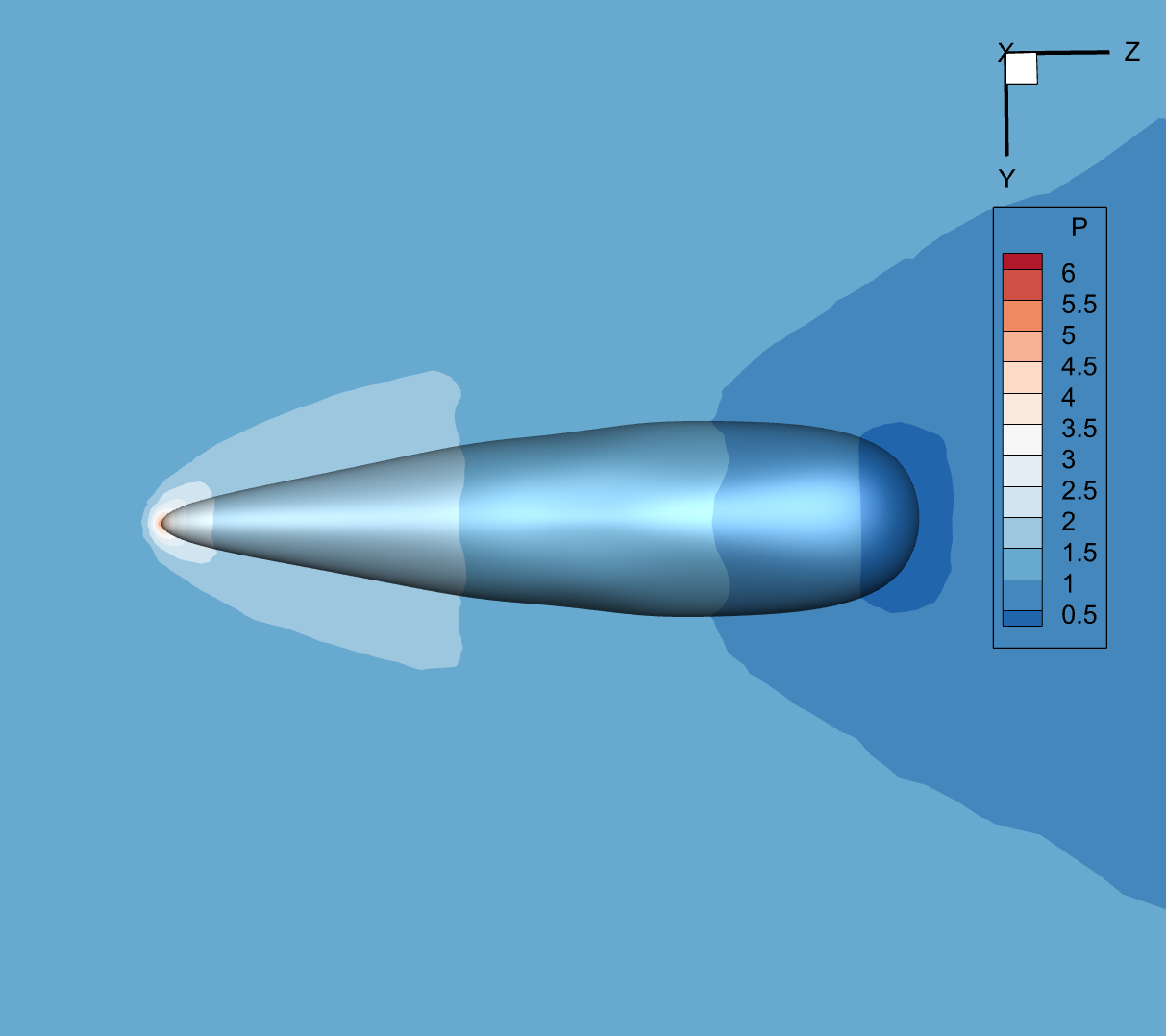}} 
    {\includegraphics[width=0.3\linewidth]{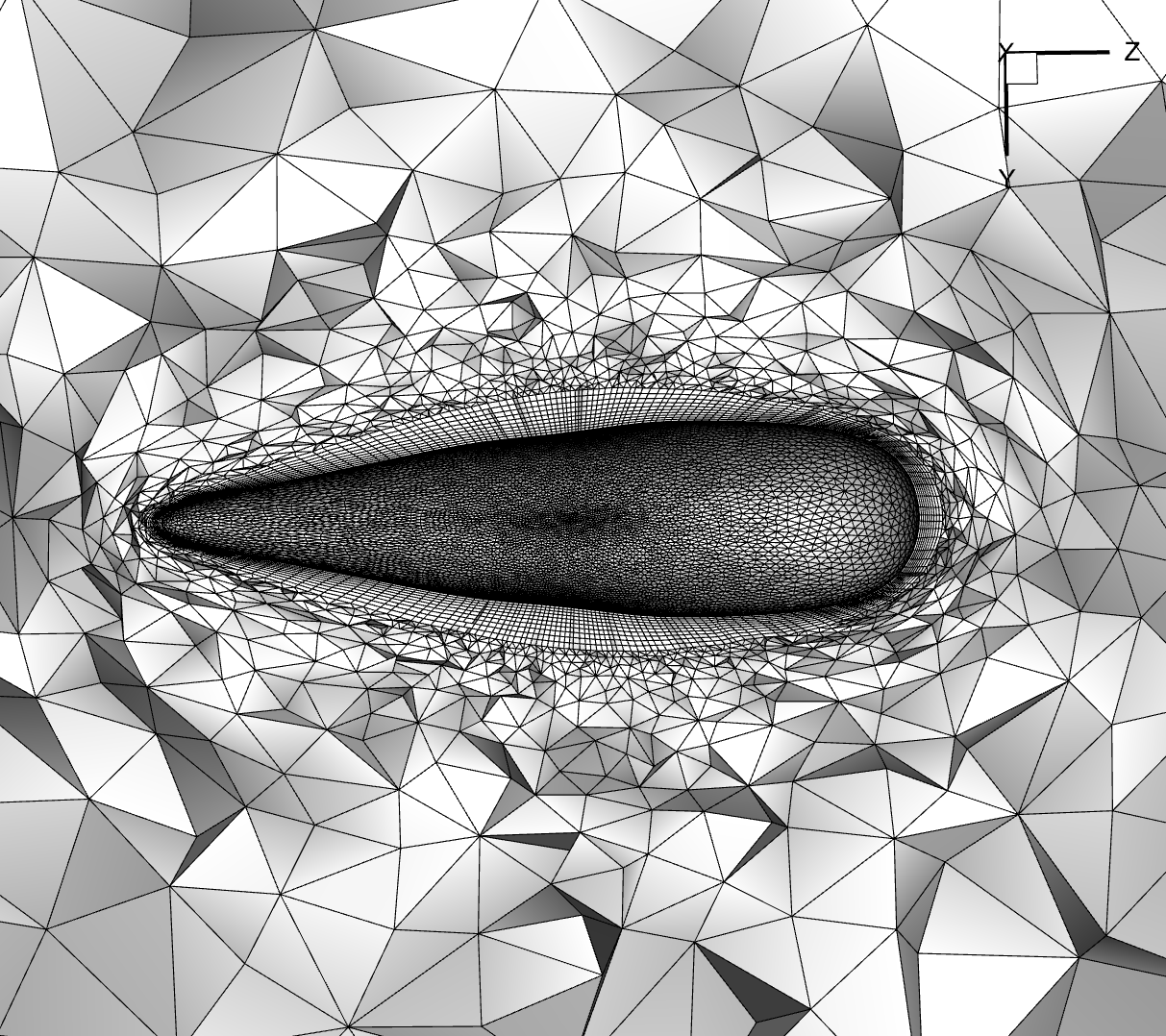}} 
    \caption{Evolution of the body shape and corresponding pressure field during optimization at $\mathrm{Kn}=0.01$ and $\mathrm{Ma}=2$. From left to right and from top to bottom, the panels show the initial shape, the shapes after the first to sixth optimization iterations, and the final optimized shape together with the deformation of the computational grid at the end of optimization in the $y$-$z$ plane with $x=0$.
    }
    \label{fig:Sphere_Ma2_Kn0d01_evolution}
\end{figure}

Fig.~\ref{fig:Sphere_Ma2_Op_Step} shows the convergence history of the dimensionless drag $F_d$. 
For both Knudsen numbers, the drag decreases rapidly during the first few iterations and then approaches a nearly constant value. 
At $\text{Kn}=0.1$, $F_d$ drops from an initial value of about $4.6$ to approximately $3.1$ after six shape adjustments, while at $\text{Kn}=0.01$ it decreases from about $3.5$ to roughly $1.8$ after five shape adjustments, corresponding to drag reductions of $34.5\%$ and $61.1\%$, respectively.

Figs.~\ref{fig:Sphere_Ma2_Kn0d1_evolution} and~\ref{fig:Sphere_Ma2_Kn0d01_evolution} illustrate the evolution of the geometry and the corresponding pressure field for $\text{Kn}=0.1$ and $\text{Kn}=0.01$, respectively. Starting from the initial spherical configuration, the optimizer quickly deforms the body into a streamlined shape that is elongated in the streamwise ($z$) direction and exhibits a blunter nose and a tapered aft region. As the iterations proceed, the high-pressure stagnation region in front of the body shrinks and its peak level is reduced, while the low-pressure wake becomes narrower and more attached to the surface. These changes are more pronounced in the slip-regime case $\text{Kn}=0.01$, where the optimized shape resembles a classical aerodynamic body of revolution~\cite{Hoerner1965}.

The final subfigures in Figs.~\ref{fig:Sphere_Ma2_Kn0d1_evolution} and~\ref{fig:Sphere_Ma2_Kn0d01_evolution} present the deformed computational grids at the end of the optimization. Despite the substantial geometric changes from the initial sphere to the optimized streamlined bodies, the near–wall prism layers remain well aligned with the surface, and the surrounding tetrahedral elements do not exhibit noticeable skewness or inversion. This demonstrates that the combination of FFD-based parameterization and the wall-distance-weighted spring smoothing~\eqref{eq:spring-kij} is able to accommodate large shape deformations while maintaining good mesh quality, which is essential for the robustness of the GSIS solver and the reliability of the computed sensitivities. 

For the two considered Knudsen numbers, the simulations were performed on an Intel Xeon Gold 6148 system using 200 and 320 cores, requiring 0.86 and 1.49 hours, respectively.  

\section{Conclusions}\label{sec:conclusion}

In summary, we have developed a fast-converging and asymptotic-preserving adjoint-based optimization
framework for three-dimensional rarefied gas flows. By extending the GSIS to the adjoint kinetic equation, the proposed method overcomes the severe stiffness of the Boltzmann model in the near-continuum regime and dramatically accelerates convergence across the entire Knudsen number spectrum. The resulting solver requires only a few dozen updates of the velocity distribution function to reach steady state, offering orders-of-magnitude improvements over conventional discrete-velocity methods.

The adjoint formulation was combined with accurate boundary sensitivities, free-form deformation  parameterization, and wall-distance-weighted spring smoothing to enable robust shape optimization on large three-dimensional meshes. Numerical examples on drag minimization of a sphere demonstrated excellent agreement between adjoint-based sensitivities and finite-difference results, rapid convergence of the adjoint GSIS solver, and substantial drag reductions, typically within ten optimization iterations. These results confirm the method's high efficiency, robustness, and suitability for complex three-dimensional design problems.

Overall, the proposed framework provides a powerful and computationally affordable tool for rarefied gas dynamics optimization, enabling practical design studies for applications such as atmospheric re-entry vehicles, vacuum systems, and micro-electromechanical devices. 


\section*{Declaration of competing interest}
The authors declare that they have no known competing financial interests or personal relationships that could have appeared to
influence the work reported in this paper.

\section*{Acknowledgments}
This work is supported by the National Natural Science Foundation of China (12402388). The authors acknowledge the computing resources from the Center for Computational Science and Engineering at the Southern University of Science and Technology.

\appendix
\section{Flux calculation for the adjoint macroscopic equations}\label{appendix_flux_adjoint}

The adjoint convective flux $\hat{\bm{F}}_{ij}^{c}$ and viscous flux $\hat{\bm{F}}_{ij}^{v} $ at the interface $ij$ are
\begin{equation}
\begin{aligned}
    \hat{\bm{F}}_{ij}^{c} & = \left( \begin{array}{c}
    u_n \hat{W}_{1} + \frac{|\bm{u}|^{2}}{2}(\gamma-1) l_{1} - u_n l_{2} \\
    u_n \hat{W}_{2} - u_{1}(\gamma-1) l_{1} + n_{1} l_{2} \\
    u_n \hat{W}_{3} - u_{2}(\gamma-1) l_{1} + n_{2} l_{2} \\
    u_n \hat{W}_{4} - u_{3}(\gamma-1) l_{1} + n_{3} l_{2} \\
    u_n \hat{W}_{5} + (\gamma-1) l_{1}
    \end{array} \right) + \frac{1}{2} \varrho_{ij}^{i} \left(\hat{\bm{W}}_j - \hat{\bm{W}}_i \right),\\
\hat{\bm{F}}_{ij}^{v} &= \frac{1}{\rho} \left( \begin{array}{c}
    -u_{I} \hat{\sigma}_{IJ} n_{J} + \left( \frac{u^{2}}{2} - \frac{P}{(\gamma-1)\rho} \right) \hat{q} \\
    \hat{\sigma}_{1J} n_{J} - u_{1} \hat{q} \\
    \hat{\sigma}_{2J} n_{J} - u_{2} \hat{q} \\
    \hat{\sigma}_{3J} n_{J} - u_{3} \hat{q} \\
    \hat{q}
    \end{array} \right),
\end{aligned}
\end{equation}
where $\hat{\bm{W}}=(\hat{W}_1,\hat{W}_2,\hat{W}_3,\hat{W}_4,\hat{W}_5)$ is the adjoint macroscopic quantity, the subscripts $i, j$ shows the cell index, and 
\begin{equation}
    \begin{aligned}
        u_n &= \bm{u} \cdot \bm{n},\quad
        l_{1} = (\bm{\varphi} \cdot \bm{n}) + u_n \hat{W}_{5}, \quad
        l_{2} = \hat{W}_{1} + (\bm{\varphi}  \cdot \bm{u}) + H \hat{W}_{5}, \\
        \varrho_{ij}^{i} &= \left| u_n \right| + \sqrt{\gamma RT} + 2 \frac{\mu_{i}}{\rho_{i} \Delta x_{ij}}, \quad 
        \hat{q} = \frac{\gamma \mu}{\text{Pr}} \nabla \hat{W}_{5} \cdot \bm{n},\\
        \hat{\sigma}_{IJ} &= \mu \left( \partial_{J}\varphi_{I} + \partial_{I}\varphi_{J} - \frac{2}{3}\delta_{IJ} \nabla \cdot \bm{\varphi} \right) + \mu \left( u_{J}\partial_{I}\hat{W}_{5} + u_{I}\partial_{J}\hat{W}_{5} - \frac{2}{3}\delta_{IJ} \bm{u} \cdot \nabla \hat{W}_{5} \right).
    \end{aligned}
\end{equation}
In the above equations, $\delta_{IJ}$ is the Kronecker delta and \(\bm{\varphi} = (\hat{W}_2, \hat{W}_3, \hat{W}_4)\). The adjoint variable \(\hat{\bm{W}}\) in \(\hat{\bm{F}}_{ij}^{c}\) is calculated as \(\hat{\bm{W}} = \frac{1}{2}(\hat{\bm{W}}_i - \hat{\bm{W}}_j)\). The gradients of the adjoint variables in $\hat{\bm{F}}_{ij}^{v}$ are  averaged at the cell face \(ij\), and \(H = E + p / \rho\) represents the total specific enthalpy. 
It should be noted that the primitive macroscopic variable $\bm{W}$ is defined in the grid cell $i$.

\section{Boundary condition for adjoint macroscopic equation}\label{appendix_B_ajoint}


\subsection{Wall boundary condition} \label{appendix_B_ajointNS_wall}

The increment of macroscopic flux at the wall for the adjoint macroscopic equations is
\begin{equation}
    \begin{aligned}
        \Delta \hat{\bm{F}}(\Delta \hat{\bm{W}}) &= \int_{\Xi ^ - }(\bm{v} \cdot \bm{n})\Delta \phi_{eq} \frac{\partial g}{\partial \bm{W}} d \Xi + \int_{\Xi ^ +}(\bm{v} \cdot \bm{n}) A_w \frac{\partial g}{\partial \bm{W}} d \Xi \\
        \\
        &= \textbf{Q}\left[\textbf{G}_{loc}\left(\textbf{K}_{loc}^- \Delta\hat{\bm{W}}_{loc} + A_w\bm{J}_{loc}^+ \right) \right].
    \end{aligned}
\end{equation}
where the local adjoint macroscopic variables $\Delta \hat{\bm{W}}_{loc} = \textbf{Q}^\top \Delta \hat{\bm{W}}$ and $\Delta \phi_{eq} = \Delta \hat{\bm{W}} \cdot \bm{\psi}$

$\textbf{Q}$ is the transformation matrix from the local coordinate system with the normal direction to the global coordinate system:
\begin{equation}
    \mathbf{Q} = 
        \left(\begin{array}{ccccc}
        1 & 0 & 0 & 0 & 0 \\
        0 & n_{x} & t_{1x} & t_{2x} & 0 \\
        0 & n_{y} & t_{1y} & t_{2y} & 0 \\
        0 & n_{z} & t_{1z} & t_{2z} & 0 \\
        0 & 0 & 0 & 0 & 1
    \end{array}\right).
\end{equation}
For the other two tangential unit direction vectors \(\bm{t}_1\) and \(\bm{t}_2\), they can be chosen arbitrarily, provided all three direction vectors are pairwise orthogonal.

Using $\textbf{Q}$, the macroscopic flow variables are transformed into the local coordinate system as:
\begin{equation}
    \begin{aligned}
        \bm{W}_{loc} &= \textbf{Q}^\top \bm{W} = (\rho, \rho u_n, \rho u_{t1},\rho u_{t2}, \rho E)^\top = (\rho, \rho \bm{u}_{loc}, \rho E)^\top.
    \end{aligned}
\end{equation}

For later convenience, we introduce the following auxiliary quantities
in the local frame:
\begin{equation}
    \begin{aligned}
        a = \sqrt{RT}, \quad u_t^2 =u_{t1}^2+u_{t2}^2,\quad \xi = \dfrac{u_{n}}{\sqrt{2}a},\quad \Phi^{-} = \frac{\operatorname{erfc}(\xi)}{2} , \quad \Phi^{+} = 1 - \Phi^{-}, \quad \Psi = \frac{\exp(-\xi^{2})}{\sqrt{2\pi}} .
    \end{aligned}
\end{equation}

$\textbf{G}$ is the coefficient matrix for the expansion of $\frac{\partial g}{\partial \bm{W}}$ in terms of $\bm{\psi}$, i.e., \( \frac{\partial g}{\partial W_i} = \psi_j g G_{ij} \). The explicit expression for $\textbf{G}_{loc}$ is:
\begin{equation}
    \mathbf{G}_{loc} = \left(\begin{array}{ccccc}
        \dfrac{15 R^{2} T^{2}+u^{2}}{6 R^{2} \rho T^{2}} & -\dfrac{u_{n} u^{2}}{3 R^{2} \rho T^{2}} & -\dfrac{u_{t 1} u^{2}}{3 R^{2} \rho T^{2}} & -\dfrac{u_{t 2} u^{2}}{3 R^{2} \rho T^{2}} & \dfrac{u^{2}-3 R T}{3 R^{2} \rho T^{2}} \\[2ex]
        -\dfrac{u_{n} u^{2}}{3 R^{2} \rho T^{2}} & \dfrac{3 R T+2 u_{n}^{2}}{3 R^{2} \rho T^{2}} & \dfrac{2 u_{n} u_{t 1}}{3 R^{2} \rho T^{2}} & \dfrac{2 u_{n} u_{t 2}}{3 R^{2} \rho T^{2}} & -\dfrac{2 u_{n}}{3 R^{2} \rho T^{2}} \\[2ex]
        -\dfrac{u_{t 1} u^{2}}{3 R^{2} \rho T^{2}} & \dfrac{2 u_{t 1} u_{n}}{3 R^{2} \rho T^{2}} & \dfrac{3 R T+2 u_{t 1}^{2}}{3 R^{2} \rho T^{2}} & \dfrac{2 u_{t 1} u_{t 2}}{3 R^{2} \rho T^{2}} & -\dfrac{2 u_{t 1}}{3 R^{2} \rho T^{2}} \\[2ex]
        -\dfrac{u_{t 2} u^{2}}{3 R^{2} \rho T^{2}} & \dfrac{2 u_{t 2} u_{n}}{3 R^{2} \rho T^{2}} & \dfrac{2 u_{t 2} u_{t 1}}{3 R^{2} \rho T^{2}} & \dfrac{3 R T+2 u_{t 2}^{2}}{3 R^{2} \rho T^{2}} & -\dfrac{2 u_{t 2}}{3 R^{2} \rho T^{2}} \\[2ex]
        \dfrac{u^{2}-3 R T}{3 R^{2} \rho T^{2}} & -\dfrac{2 u_{n}}{3 R^{2} \rho T^{2}} & -\dfrac{2 u_{t 1}}{3 R^{2} \rho T^{2}} & -\dfrac{2 u_{t 2}}{3 R^{2} \rho T^{2}} & \dfrac{2}{3 R^{2} \rho T^{2}}
    \end{array}\right).
\end{equation}

The specific forms of all components of the symmetric local half-space matrix $\textbf{K}_{loc}^-$ are:
\begin{equation}
    \begin{aligned}
        K_{11}^{-}&=I_1^{-}, \quad
        K_{12}^{-}=I_2^{-},\quad
        K_{13}^{-}=u_{t1}I_1^{-},\quad
        K_{14}^{-}=u_{t2}I_1^{-},\quad
        K_{15}^{-}=\tfrac12\!\left[I_3^{-}+(u_t^2+2a^2)I_1^{-}\right],\\[2pt]
        K_{22}^{-}&=I_3^{-},\quad
        K_{23}^{-}=u_{t1}I_2^{-},\quad
        K_{24}^{-}=u_{t2}I_2^{-},\quad
        K_{25}^{-}=\tfrac12\!\left(I_4^{-}+(u_t^2+2a^2)I_2^{-}\right),\\[2pt]
        K_{33}^{-}&=(u_{t1}^2+a^2)I_1^{-},\quad
        K_{34}^{-}=(u_{t1}u_{t2})I_1^{-},\quad
        K_{35}^{-}=\tfrac12\!\left(u_{t1}I_3^{-}+u_{t1}(u_t^2+4a^2)I_1^{-}\right),\\[2pt]
        K_{44}^{-}&=(u_{t2}^2+a^2)I_1^{-},\quad
        K_{45}^{-}=\tfrac12\!\left(u_{t2}I_3^{-}+u_{t2}(u_t^2+4a^2)I_1^{-}\right),\\[2pt]
        K_{55}^{-}&=\tfrac14\!\left(I_5^{-}+2(u_t^2+2a^2)I_3^{-}
         +(u_t^4+8a^2u_t^2+8a^4)I_1^{-}\right).
    \end{aligned}
\end{equation}
Here, the scalar half-space integrals $I_m^-$ with $m=1,2,3,4,5$ are given by:
\begin{equation}
    \begin{aligned}
        I_1^- &= \rho\left(u_n\Phi^- - a\,\Psi\right),\\
        I_2^- &= \rho\left((u_n^2+a^2)\Phi^- - a\,u_n\,\Psi\right),\\
        I_3^- &= \rho\left((u_n^3+3a^2u_n)\Phi^- - a\,(u_n^2+2a^2)\,\Psi\right),\\
        I_4^- &= \rho\left((u_n^4+6a^2u_n^2+3a^4)\Phi^- - a\,(u_n^3+5a^2u_n)\,\Psi\right),\\
        I_5^- &= \rho\left((u_n^5+10a^2u_n^3+15a^4u_n)\Phi^- - a\,(u_n^4+9a^2u_n^2+8a^4)\,\Psi\right).
    \end{aligned}
\end{equation}

The local incident half-space moment vector is written as:
\begin{equation}
    \begin{aligned}
    \bm{J}_{loc}^+ &= \begin{pmatrix}
        J_1^{+},
        (J_{\bm{v}}^{+})_{n},
        (J_{\bm{v}}^{+})_{t1},
        (J_{\bm{v}}^{+})_{t2},
        J_5^{+}
    \end{pmatrix}^\top,\\
        J_1^{+}&=\rho\,(u_n\Phi^+ + a\Psi),\\
        J_{nn}^{+}&=\rho\big((u_n^2+a^2)\Phi^++a u_n\Psi\big),\\
        J_3^{+}&=\rho\big((u_n^3+3u_n a^2)\Phi^++a(u_n^2+2a^2)\Psi\big),\\
        J_{\bm{v}}^{+}&=\bm{u}_{loc} J_1^{+}+\bm{e}_1\big(J_{nn}^{+}-u_nJ_1^{+}\big),\\
        J_5^{+}&=\tfrac12\big(J_3^{+}+(u_t^2+2a^2)J_1^{+}\big).
    \end{aligned}
\end{equation}
where $\bm{e}_1=(1, 0, 0)$.

Finally, the wall coefficient $A_w$ for an isothermal stationary wall, with $g_w = g_{\rm M}(1,\bm{0},T_w)$, is obtained as:
\begin{equation}
    \begin{aligned}
        A_w = -\sqrt{\frac{2 \pi}{RT_w}} \int_{\Xi ^ - } (\bm{v} \cdot \bm{n}) \Delta \phi_{eq} g_w d \Xi 
        = \Delta \hat{W}_{1} - \sqrt{\frac{\pi RT_w}{2}} \left( \Delta \bm{\varphi}  \cdot \boldsymbol{n} \right) + 2RT_w \Delta \hat{W}_{5}.
    \end{aligned}
\end{equation}


\subsection{Dirichlet boundary condition}

The increment of macroscopic flux at the Dirichlet boundary (e.g., far-field) for the adjoint macroscopic equations is:
\begin{equation}
    \begin{aligned}
        \Delta \hat{\bm{F}}(\Delta \hat{\bm{W}}) &= \int_{\Xi ^ - }(\bm{v} \cdot \bm{n})\Delta \phi_{eq} \frac{\partial g}{\partial \bm{W}} d \Xi
        = \textbf{Q}\textbf{G}_{loc}\left(\textbf{K}_{loc}^- \Delta \hat{\bm{W}}_{loc} \right).
    \end{aligned}
\end{equation}
The calculation of this equation can be found in \ref{appendix_B_ajointNS_wall}.

\bibliographystyle{elsarticle-num}

\bibliography{ref}

\begin{thebibliography}{10}
\expandafter\ifx\csname url\endcsname\relax
  \def\url#1{\texttt{#1}}\fi
\expandafter\ifx\csname urlprefix\endcsname\relax\def\urlprefix{URL }\fi
\expandafter\ifx\csname href\endcsname\relax
  \def\href#1#2{#2} \def\path#1{#1}\fi

\bibitem{reed2010investigation}
E.~Reed, H.~Alkandry, J.~Codoni, J.~McDaniel, I.~Boyd, Investigation of the
  interactions of reaction control systems with mars science laboratory
  aeroshell, in: 48th AIAA Aerospace Sciences Meeting \& Exhibit, 2010, p.
  1558.

\bibitem{li2021kinetic}
J.~Li, D.~Jiang, X.~Geng, J.~Chen, Kinetic comparative study on aerodynamic
  characteristics of hypersonic reentry vehicle from near-continuous flow to
  free molecular flow, Advances in Aerodynamics 3 (2021) 1--10.

\bibitem{hablanian1997high}
M.~H. Hablanian, High-vacuum technology: a practical guide, CRC Press, 1997.

\bibitem{sharipov2005numerical}
F.~Sharipov, P.~Fahrenbach, A.~Zipp, {Numerical modeling of the Holweck pump},
  Journal of Vacuum Science \& Technology A 23 (2005) 1331--1339.

\bibitem{bakshi2009euv}
V.~Bakshi, {EUV lithography}, SPIE press, 2009.

\bibitem{tantos2020deterministic}
C.~Tantos, S.~Varoutis, C.~Day, Deterministic and stochastic modeling of
  rarefied gas flows in fusion particle exhaust systems, Journal of Vacuum
  Science \& Technology. B. 38 (2020) 064201.

\bibitem{leifsson2015aerodynamic}
L.~Leifsson, S.~Koziel, Aerodynamic shape optimization by variable-fidelity
  computational fluid dynamics models: a review of recent progress, Journal of
  Computational Science 10 (2015) 45--54.

\bibitem{jameson1988aerodynamic}
A.~Jameson, Aerodynamic design via control theory, Journal of Scientific
  Computing 3 (1988) 233--260.

\bibitem{han2012hierarchical}
Z.-H. Han, S.~G{\"o}rtz, Hierarchical {Kriging} model for variable-fidelity
  surrogate modeling, AIAA Journal 50 (2012) 1885--1896.

\bibitem{sun2019review}
G.~Sun, S.~Wang, A review of the artificial neural network surrogate modeling
  in aerodynamic design, Proceedings of the Institution of Mechanical
  Engineers, Part G: Journal of Aerospace Engineering 233~(16) (2019)
  5863--5872.

\bibitem{Yang1995Rarefied}
J.~Y. Yang, J.~C. Huang, Rarefied flow computations using nonlinear model
  {Boltzmann} equations, Journal of Computational Physics 120~(2) (1995)
  323--339.

\bibitem{mieussens2000}
L.~Mieussens, Discrete velocity model and implicit scheme for the {BGK}
  equation of rarefied gas dynamics, Mathematical Models and Methods in Applied
  Sciences 10~(08) (2000) 1121--1149.

\bibitem{bird1994molecular}
G.~A. Bird, {Molecular Gas Dynamics and the Direct Simulation of Gas Flows},
  Oxford Science Publications, Oxford University Press Inc, New York, 1994.

\bibitem{sato2019topology}
A.~Sato, T.~Yamada, K.~Izui, S.~Nishiwaki, S.~Takata, A topology optimization
  method in rarefied gas flow problems using the {Boltzmann} equation, Journal
  of Computational Physics 395 (2019) 60--84.

\bibitem{guan2023densityIPDSMC}
K.~Guan, K.~Matsushima, Y.~Noguchi, T.~Yamada, Topology optimization for
  rarefied gas flow problems using density method and adjoint {IP-DSMC},
  Journal of Computational Physics 474 (2023) 111788.

\bibitem{guan2024adjointDVM}
K.~Guan, T.~Yamada, Topology optimization of rarefied gas flows using an
  adjoint discrete velocity method, Journal of Computational Physics 511 (2024)
  113111.

\bibitem{yuan2024design}
R.~Yuan, L.~Wu, A design optimization method for rarefied and continuum gas
  flows, Journal of Computational Physics 517 (2024) 113366.

\bibitem{yuan2025adjoint}
R.~Yuan, L.~Wu, Adjoint shape optimization from the continuum to free-molecular
  gas flows, Journal of Computational Physics (2025) 114102.

\bibitem{Bhatnagar1954}
P.~L. Bhatnagar, E.~P. Gross, M.~Krook, A model for collision processes in
  gases. {I}. {S}mall amplitude processes in charged and neutral one-component
  systems, Phys. Rev. 94 (1954) 511--525.

\bibitem{su2020can}
W.~Su, L.~Zhu, P.~Wang, Y.~Zhang, L.~Wu, Can we find steady-state solutions to
  multiscale rarefied gas flows within dozens of iterations?, Journal of
  Computational Physics 407 (2020) 109245.

\bibitem{su2020fast}
W.~Su, L.~Zhu, L.~Wu, Fast convergence and asymptotic preserving of the general
  synthetic iterative scheme, SIAM Journal on Scientific Computing 42~(1)
  (2020) B1517--B1544.

\bibitem{zhang2024efficient}
Y.~Zhang, J.~Zeng, R.~Yuan, W.~Liu, Q.~Li, L.~Wu, Efficient parallel solver for
  rarefied gas flow using {GSIS}, Computers \& Fluids 281 (2024) 106374.

\bibitem{Cercignanibook1988}
C.~Cercignani, {The Boltzmann Equation and its Applications}, Springer-Verlag,
  New York, 1988.

\bibitem{tsien1946superaerodynamics}
H.-S. Tsien, Superaerodynamics, mechanics of rarefied gases, Journal of the
  Aeronautical Sciences 13~(12) (1946) 653--664.

\bibitem{svanberg1987method}
K.~Svanberg, The method of moving asymptotes---a new method for structural
  optimization, International Journal for Numerical Methods in Engineering
  24~(2) (1987) 359--373.

\bibitem{svanberg2002class}
K.~Svanberg, A class of globally convergent optimization methods based on
  conservative convex separable approximations, SIAM Journal on Optimization
  12~(2) (2002) 555--573.

\bibitem{johnson2007nLopt}
S.~G. Johnson, The {NLopt} nonlinear-optimization package,
  \url{https://github.com/stevengj/nlopt} (2007).

\bibitem{wang2018comparative}
P.~Wang, M.~T. Ho, L.~Wu, Z.~Guo, Y.~Zhang, A comparative study of discrete
  velocity methods for low-speed rarefied gas flows, Computers \& Fluids 161
  (2018) 33--46.

\bibitem{Rogers1995Comparison}
S.~E. Rogers, Comparison of implicit schemes for the incompressible
  {Navier-Stokes} equations, AIAA Journal 33~(11) (1995).

\bibitem{bueno2012continuous}
A.~Bueno-Orovio, C.~Castro, F.~Palacios, E.~Zuazua, Continuous adjoint approach
  for the {Spalart-Allmaras} model in aerodynamic optimization, AIAA Journal 50
  (2012) 631--646.

\bibitem{yuan2021novel}
R.~Yuan, S.~Liu, C.~Zhong, A novel multiscale discrete velocity method for
  model kinetic equations, Communications in Nonlinear Science and Numerical
  Simulation 92 (2021) 105473.

\bibitem{liu2024further}
W.~Liu, Y.~Zhang, J.~Zeng, L.~Wu, Further acceleration of multiscale simulation
  of rarefied gas flow via a generalized boundary treatment, Journal of
  Computational Physics 503 (2024) 112830.

\bibitem{sederberg1986free}
T.~W. Sederberg, S.~R. Parry, Free-form deformation of solid geometric models,
  in: Proceedings of the 13th annual conference on Computer graphics and
  interactive techniques, 1986, pp. 151--160.

\bibitem{batina1990unsteady}
J.~T. Batina, Unsteady {Euler} airfoil solutions using unstructured dynamic
  meshes, AIAA journal 28~(8) (1990) 1381--1388.

\bibitem{matsushima2002unstructured}
K.~Matsushima, M.~Murayama, K.~Nakahashi, Unstructured dynamic mesh for large
  movement and deformation, in: 40th AIAA Aerospace Sciences Meeting \&
  Exhibit, 2002, p. 122.

\bibitem{duvigneau2006multi}
R.~Duvigneau, B.~Chaigne, J.-A. D{\'e}sid{\'e}ri, Multi-level parameterization
  for shape optimization in aerodynamics and electromagnetics using a particle
  swarm optimization algorithm, Ph.D. thesis, INRIA (2006).

\bibitem{palacios2012adjoint}
F.~Palacios, J.~Alonso, M.~Colonno, J.~Hicken, T.~Lukaczyk, Adjoint-based
  method for supersonic aircraft design using equivalent area distribution, in:
  50th AIAA Aerospace Sciences Meeting, 2012, p. 269.

\bibitem{economon2016su2}
T.~D. Economon, F.~Palacios, S.~R. Copeland, T.~W. Lukaczyk, J.~J. Alonso,
  {SU2: An open-source suite for multiphysics simulation and design}, Aiaa
  Journal 54~(3) (2016) 828--846.

\bibitem{Hoerner1965}
S.~F. Hoerner, {Fluid-Dynamic Drag}, Hoerner Fluid Dynamics, 1965.

\end{thebibliography}

\end{document}